\documentstyle[aps,pre,multicol,epsf,psfig,boxedminipage,citesort,graphicx]{revtex}

\begin{document}

\newcommand{\vmax}{v_{\text{max}}}

\twocolumn[\hsize\textwidth\columnwidth\hsize\csname@twocolumnfalse%
\endcsname

\title{An empirical test for cellular automaton models of traffic flow}

\author{Wolfgang  Knospe$^{1}$,  Ludger    Santen$^{2}$,     Andreas
Schadschneider$^{3}$, Michael Schreckenberg$^{1}$ \\} 

\address{$^{1}$Theoretische Physik FB 10,
Gerhard-Mercator-Universit\"at   Duisburg,  D-47048 Duisburg, Germany}
\address{$^{2}$ Fachrichtung Theoretische Physik, Universit\"at des
Saarlandes, Postfach 151150, 66041 Saarbr\"ucken, Germany}
\address{$^{3}$ Institut  f\"ur Theoretische  Physik, Universit\"at zu
K\"oln D-50937 K\"oln, Germany} 

\date{\today} 

\maketitle 

\begin{abstract}
 Based on a detailed microscopic test scenario motivated by recent
 empirical studies of single-vehicle data, several cellular automaton
 models for traffic flow are compared. We find three levels of agreement 
 with the empirical data: 1) models that do not reproduce even 
 qualitatively the most important empirical observations, 
 2) models that are on a macroscopic level in reasonable agreement
 with the empirics, and 3) models that  reproduce the empirical data 
 on a microscopic level as well.
 Our results are not only relevant for applications, but also
 shed new light on the relevant interactions in traffic flow.
\end{abstract}
\bigskip
]


\section{Introduction}

For a long time the  modeling of traffic  flow phenomena was dominated
by two theoretical approaches (for a review, see 
e.g., \cite{review,Helbing2000,nagatani,NWW}).
The   first type of models,  the  so-called  car-following models, are
based on the  fact that the behavior of  a driver is determined by the
leading vehicle. This assumption leads to dynamical velocity equations
which in general depend on the distance to the leading vehicles and on
the  velocity difference  between    the  leading and   the  following
vehicle. An  alternative approach, which is   also well established in
traffic research, does not treat the  individual cars but describes the
dynamics of traffic networks in terms of macroscopic variables.  Here
traffic flow phenomena are treated in analogy to the dynamics of
compressible viscous fluids. 

Both approaches are still widely used by traffic engineers, but for
practical purposes they are often not suitable. One of the main
problems of present car-following models (e.g., see
\cite{herman,gazis61,gazis67,gipps81,rothery_trb}) is that they are
difficult to treat in computer simulations of large networks. On the
other hand also the macroscopic approaches lead to some difficulties
although large networks can be treated in principle.  First of all,
present macroscopic models use a large number of parameters which have
partly no counterpart within empirical investigations.  In addition to
that, the information that can be obtained using macroscopic models is
incomplete in the sense that it is not possible to trace individual cars.

In  order to fill  this gap cellular automaton (CA) models have been
invented~\cite{Nagel93,SSNI}. CA models are microscopic models which are by
design well suited  for large-scale computer simulations. A comparison
of the simulations with empirical data shows  that already very simple
approaches give meaningful results. In particular they  can be used in
order  to simulate dense  networks like  cities~\cite{esser} which are
controlled by the dynamics at the  intersections. For highway
traffic, however, a  more  detailed  description of   the  dynamics 
seems to  be necessary. 

In  this work we want  to discuss the realism  and the limitations of a 
number of CA models.  Our choice is restricted to models that are 
discrete in space and time, which e.g.\ excludes the approach by Krauss 
et al.\ \cite{krauss}, and have local interactions only, excluding 
models as the Galilei-invariant model introduced in \cite{galilei}.   
We compare simulations of the CA model proposed by Nagel and Schreckenberg, 
that is to date the most frequently used CA  approach  for traffic flow,  
the VDR model~\cite{robert} which realizes a so-called slow-to-start
rule, the TOCA-model of Brilon {\it et al.}~\cite{brilon}, the model
of  Emmerich and Rank \cite{emmerich} based on the use of velocity-gap
matrices, and  the  approach by  Helbing  and
Schreckenberg \cite{HeSch} which represents a  model with a  more
sophisticated distance rule. 
Finally we discuss the recently introduced brake light 
model~\cite{knospe2001,knospe2002} that was suggested in order to give 
a reliable reproduction of the microscopic empirics and the model
by Kerner, Klenov and Wolf \cite{kkw}, focusing more on the 
macroscopic properties of the three phases of traffic flow.

We will compare the ability of these models to reproduce the 
empirical findings. This requires using a measurement procedure in 
the simulations which models the detectors on the highway.
Analogous to the  empirical  setup of~\cite{neubert}  the simulation
data are evaluated by a virtual inductive loop, i.e., speed and time-headway 
of the vehicles are measured at a given  link of the
lattice. The measurement process is applied after the update of the
velocity has been carried out, but right before the movement of the
vehicles. This implies that the gap to the preceding vehicle does not
change significantly during the measurement. These simulation data are
analyzed regarding individual and  
aggregated quantities, as it has been done in recent empirical
investigations \cite{neubert,Tilch99TGF,knospe2002b}.

Although most of the empirical data sets have been collected at 
multi-lane highways, we have performed our simulations on a  single-lane
road in order to reduce the number of adjustable parameters. This 
approach is justified because the empirical data sets are selected such that 
multi-lane effects are of minor importance. They might play a role
for synchronized traffic of type $(i)$ and $(ii)$, as it has been recently
argued in  \cite{mahnke}, but in any case these types of synchronized 
traffic are much rarely observed than synchronized traffic of type $(iii)$
\cite{neubert,knospe2002b}.

We will also not consider effects by a mixture of different vehicle
types, e.g.\ there are no trucks in our simulations. The fraction of
slow cars has not been determined from the empirical data. Furthermore
these data have been collected on a highway with speed limit such
that disorder effects through slower cars are expected to play a minor
role. We believe that inclusion of disorder will not change our
results qualitatively, but can lead to a better quantitative agreement
in some cases.

Before we start the analysis of the above mentioned CA models, we will 
introduce an empirical test scenario. 
It will be microscopic and local to make it easily comparable
to online data provided e.g.\ by inductive loops. In contrast, the 
detection of complex spatio-temporal structures \cite{kernerNet} 
is more difficult to achieve in an automated way. It would require
the investigation of interface dynamics whereas in our scenario
only bulk properties are studied.
This test scenario also 
verifies the reproduction of empirical traffic states on a microscopic 
level, a task that cannot be fulfilled by macroscopic models.
 The empirical results have been 
chosen with respect to their reproducibility and the ability to distinguish 
between the different states of traffic. This scenario will be discussed 
in the next section. 


\section{Empirical facts}

In order to probe the accuracy and the degree of realism of the
different models one has to introduce a test scenario that includes
the most important empirical findings.  The difficulty in defining
such a scenario is due to the fact that the empirical results may
depend strongly on the particular environment.  Therefore one has to
try to extract the results that really characterize the behavior of
the vehicles. As an additional difficulty mostly aggregated data have
been analyzed which are known to be largely dependent on the road
conditions, e.g., the capacity of an upstream bottleneck. A number of
results, however, is of general nature as we will discuss below.

Even more conclusive are empirical investigations that use 
single-vehicle data. These measurements can be compared directly to the 
simulation results and include important information concerning the 
microscopic structure of vehicular traffic. Unfortunately only 
a small number of empirical investigations based on 
single-vehicle data exists so far. Our discussion refers to the 
empirical studies of refs.~\cite{neubert,Tilch99TGF,knospe2002b}.
In particular, in order to reduce the effects of disorder,
the results of~\cite{neubert} (except for the time-headway
distributions, see below) are used for the comparison with simulation 
data. These data have been collected on 
a highway where a speed limit applies. This facilitates the
comparison with modeling approaches.

The empirical findings that are taken as a basis for the 
comparison with the model results have been obtained from inductive 
loops. Measurements by inductive loops, which represent the most frequently 
used measurement devices, give information about the number of 
cars passing, their velocities and the occupation times. These 
direct measurements are also used in order to calculate other 
quantities, e.g., the spatial distance $d_n$ via 
$d_n = v_{n-1}  t_h$ (where $v_{n-1}$ denotes the velocity 
of the preceding car ${n-1}$, $t_h$ the time-headway between car 
$n-1$ and car $n$). 


\subsection{Temporally aggregated data}

The   most important empirical quantity  is   the relation between the
averaged  observables flow and  density, i.e.,  the fundamental diagram.
There exists  a longstanding controversy (see e.g.\ 
\cite{kernerNet,HTcoop} and references therein) about  the  ``correct''
functional form of the fundamental diagram  and  a large  number  of
possible forms  have been suggested  to be compatible with
empirical data~\cite{hall}. 
A more consistent picture was established after the work of Kerner and
coworkers who distinguished at least three different phases of traffic
flow~\cite{kernerphysworld}, i.e., free flow, synchronized traffic and
wide jams, that have to be  
analyzed separately.  We will follow this scheme
and summarize the empirical findings accordingly.

Usually these measurements are stored as averaged values of certain 
time-intervals. We discuss results for the fundamental diagram,
i.e., the flow density relation, in the different traffic phases that 
are based on one-minute data. The results for the functional form of
the flow are shown, as far as possible, in dependence of the spatial 
density $\rho(t)$. The density can be calculated from
\begin{equation} 
\rho(t) = \frac{J(t)}{v(t)}, 
\end{equation}
where  $J(t)$ denotes the number of cars passing the detector with an
average velocity $v(t)$ in the corresponding time interval.

{\em Free flow} traffic is characterized by a large value of the
average speed. One basically observes two qualitatively different
functional forms of the fundamental diagram, i.e., that the linear
regime extends up to the observed maximum of the flow or that one has
a finite curvature in particular for densities slightly below the
density of maximum flow \cite{knospe2002b,NeubertDiss}.  The finite
curvature is a consequence of an alignment of speeds, i.e., close to
the optimal flow it is not longer possible to drive systematically
faster than the trucks. This point of view is supported by the
empirical results taken from highways where a quite restrictive speed
limit is applied that can be reached even by trucks \cite{neubert}.

In this case the whole free flow branch is linear.  
For our purposes the linear form of the fundamental diagram is
relevant, because we use a single type of cars in the simulations,
with a maximal velocity that is given by the slope of the free flow
branch. When simulating a section of the highway where no speed limit
is applied, one has to take a distribution of maximal speeds. This
distribution can be obtained from the empirical velocity distributions
at very low densities, where interactions between cars can be
neglected.

In the {\em congested regime} one distinguishes between synchronized
traffic and wide jams. In the {\em synchronized phase}, the mean
velocity of the vehicles is reduced, compared to the free flow, but
the flow can take on values close to the maximum flow.  Moreover,
strong correlations between the density on different lanes exist
caused by lane changings.

\begin{figure}[hbt]
\begin{center}
\includegraphics[width=0.9\linewidth]{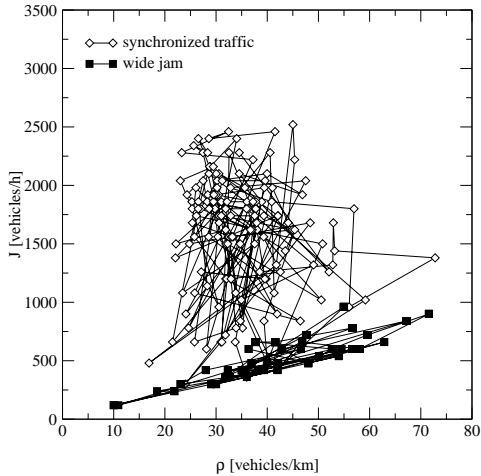}
\caption{Time-traced fundamental diagram of the two congested states
(from~[30]). Synchronized traffic is characterized by 
strong fluctuations of the density and flow. The measurements for 
wide jams are similar to measurements in free flow but with
much smaller average velocity.}
\label{fig:emp_fund}
\end{center}
\end{figure}

 The synchronized 
state has been subdivided into three types, which differ in the 
characteristics of the time series of density and flow:
In  synchronized traffic of  type $(i)$  constant  values of the density and
the flow can be  observed during a long  period  of time.  In synchronized
traffic of   type  $(ii)$ the flow   depends  linearly on   the  density
similarly to free flow, but the mean velocity is reduced considerably.
In synchronized traffic  of type $(iii)$ irregular  patterns of flow and
density can be observed (Fig.~\ref{fig:emp_fund}). In our article we
concentrate on synchronized  
flow of type $(iii)$, because the two other types of synchronized traffic 
have been rarely observed and it is not confirmed whether they are generic 
phases of traffic flow. An identification of synchronized traffic by means 
of the fundamental diagram may be misleading, because the results often 
depend on the averaging procedure. A more sensitive check is  to identify
 the  different  types of traffic states by
means of the cross-correlation  $cc(\rho,J)$ of the density $\rho$ and
the flow $J$~\cite{neubert}: 
\begin{equation} 
cc(\rho,J)    = \frac{\langle  \rho(t)J(t+\tau)\rangle - \langle
\rho(t) \rangle  \langle J(t+\tau)\rangle}
{\sqrt{\Delta \rho(t)}\ \sqrt{\Delta J(t+\tau)}}
\label{crossdef}
\end{equation}
with $\Delta A = \langle A^2 \rangle-\langle A \rangle^2$ denoting the 
variance of the observable $A$. 
The linear dependency of  the flow  and  the density in the  free flow
state as well as in the wide jam  state  leads  to
cross-correlations of $\approx 1$,  whereas irregular patterns of  the
flow and the density in the synchronized traffic of type $(iii)$ lead to
cross-correlations of $\approx 0$. 

It is worth pointing out that the notion of ``synchronized traffic''
is still very controversial \cite{HTcoop,kernerNet}. We emphasize
here, that we use an {\it objective} criterion, namely the vanishing
of the cross-correlation function~(\ref{crossdef}) for the
classification. Within the empirical single-vehicle data sets available
the other two synchronized states could not be clearly identified. 
Therefore it was not reasonable to include these states into the test 
scenario. The characteristics of synchronized traffic of type $(iii)$,
however, have been clearly distinguished from free flow and jammed states by
the criterion $cc(\rho,J) \approx 0$. 
Therefore any detailed model should be able to reproduce this class
of synchronized states.

Fig.~\ref{fig:emp_fund} includes a typical measurement of the 
fundamental diagram that correspond to {\em wide jams}. Surprisingly 
these measurements reveal quite
small values of the density, although the road is almost completely 
covered by cars. This seemingly incorrect result is due to the  
local nature of the measurement (see \cite{neubert} for a detailed 
discussion). Thus, the form of the fundamental diagram in the jammed 
state is similar to free flow traffic, but with a small average velocity.

The jammed branch of the fundamental diagram is often not
 reproduced by CA models, because they use the inverse density of 
a jam in order to calibrate the unit of length. Within these 
approaches jams are compact. In this case  (almost) no internal flux is
observed. The modeling of jams can however be meaningful, if 
the upstream velocity and other macroscopic characteristics of
jam are reproduced.


\subsection{Single vehicle data}

Nowadays some empirical studies exist that have analyzed single-vehicle 
data from counting loops \cite{neubert,Tilch99TGF,knospe2002b}. 
These studies are of great importance for 
the modeling of traffic flow because they give direct information
about the ``microscopic structure'' of traffic streams. The data usually 
include direct measurements of the time-headways and the velocities
of the vehicles as well as the occupation time of the detector.
Similar to the time-averaged observables the results for the microscopic 
quantities differ qualitatively  in the different phases.

\begin{figure}[hbt]
\begin{center}
\includegraphics[width=0.9\linewidth]{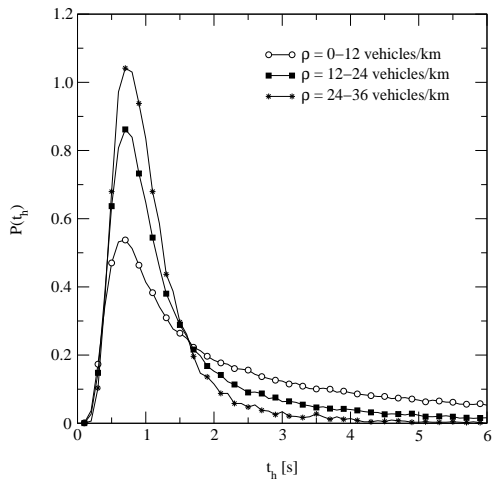}
\caption{Empirical time-headway distributions, i.e., the relative 
frequency of a given time-headway, in free flow 
traffic. The distributions are normalized, i.e., $\sum P(t_h) \cdot
\Delta t_h = 1$. The data are classified 
in different density regimes by the corresponding one-minute
data of the density. 
For a given road section one obtains a maximum that is independent of
the density and a minimal headway of 0.2~s. } 
\label{fig:emp_th_ff}
\end{center}
\end{figure}

The first quantity we look at is the time-headway distribution~\footnote{Since
the time-headway distribution of~\cite{neubert} in free flow as well as 
in the synchronized state shows some peculiarities due to an error of the 
measurement software \cite{knospe2002b}, new measurements at the same 
location have been conducted.}, i.e., the time elapsing 
between two cars passing the detector. This quantity 
is the microscopic analogue to the inverse flow.  
In free flow traffic one has found that the distribution 
at short times and also the position of the maximum is independent 
of the density (Fig.~\ref{fig:emp_th_ff}).

The {\em cut-off} at small time-headways as 
well as the {\em typical} time-headway in free flow traffic are important 
observables which have to be reproduced by the microscopic models. 
The exact shape of the distribution may also depend on the relative 
frequency of slow vehicles, because this determines the fraction of 
interacting vehicles at a given density.

\begin{figure}[hbt]
\begin{center}
\includegraphics[width=0.9\linewidth]{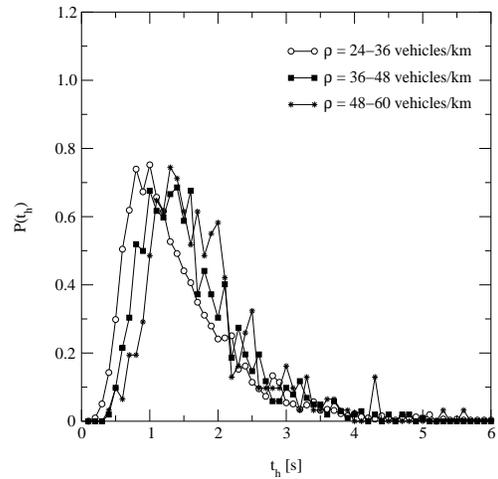}
\caption{Same as Fig.~{\ref{fig:emp_th_ff}} but for synchronized 
traffic. The functional behavior of the distribution at short times
depends on the density.}
\label{fig:emp_th_sync}
\end{center}
\end{figure}

The time-headway distributions in synchronized
traffic~\footnote{Unfortunately, new measurements taken from the
detector location used in~\cite{neubert}
do not provide a sufficient amount of data of the synchronized
state. Since the time-headway distributions
of~\cite{neubert} cannot be used, the distribution is
calculated from data sets taken from~\cite{knospe2002b}. This is
justified because the effects of 
a speed limit can be neglected at larger densities.} differ 
systematically from the free flow distributions
(Fig.~\ref{fig:emp_th_sync}). In synchronized  
traffic the distributions  have a maximum that is much broader 
than that in free flow traffic. The maximum is less pronounced 
and its position depends significantly on the density. 

In the presence of wide jams one has to distinguish between 
the jam itself and its outflow region. In the jam one finds 
evidently a broad distribution of time-headways, because cars 
are blocked for quite long times. In the outflow region of 
a jam,  however, one observes that the typical time-headway 
is of the order of $\sim 2$ s.

The characteristics of traffic jams are one of the extensively studied 
phenomena in traffic flow. Wide traffic jams can be  identified by a 
sharp drop of the velocity and the flow to  negligible
values in the time-series. Moreover traffic jams move upstream with a
surprisingly constant  velocity (typically $15$ km/h~\cite{kerner_prl81}). 
The upstream velocity is intimately related to the outflow $J_{\rm{out}}$ 
from a jam which also takes on constant values for a given situation. 
This allows the observed coexistence of jams.
The coexistence is facilitated because the outflow from a jam is considerably 
smaller than the maximal flow $J_{\rm{max}}$, such that no new jams emerge in 
the outflow region of a jam. Empirically one observes the ratio  
$J_{\rm{max}}/J_{\rm{out}} \approx  1.5$ \cite{kerner96}.  
The outflow and the upstream velocity of a jam can therefore also be 
used to calibrate the model. The precise data for the average upstream 
velocities and $J_{\rm{out}}$ may also serve 
to evaluate the average space $l$ that is occupied by a car in
a jam. Usually $l$, and not the average length of the 
vehicles, represents the length of a cell in the CA models. $l$   
may also be used to assign a reasonable value of the velocity of 
cars in a jam, i.e.,  $v_n = l /  t_h$.

\begin{figure}[hbt]
\begin{center}
\includegraphics[width=0.9\linewidth]{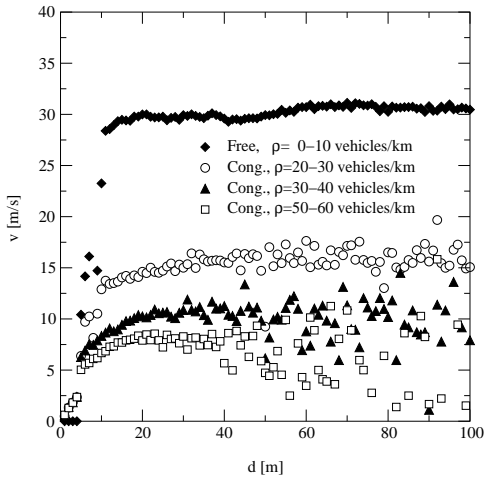} 
\caption{Empirical optimal-velocity (OV) functions, i.e., 
speed-distance relations. The figure shows the mean velocity 
for a given spatial distance in free flow and congested traffic for
different densities.}  
\label{ov_emp}
\end{center}
\end{figure}

The final test of the models comes from the velocity distance relation
in the different traffic phases (Fig.~\ref{ov_emp}). This relation,
also called optimal-velocity (OV) function, characterizes in great 
detail the microscopic structure of the different phases. Some models
use OV-curves directly as 
an input~\cite{bando}. In any case this quantity is a sensitive test 
concerning the reproduction of the microscopic structure of highway
traffic. In the free   flow   regime the  asymptotic velocity does not
depend on the density, but is given by the applied speed limit.  
In the congested regime  this  asymptotic velocity is much
smaller  than  in free flow, i.e., cars  are  driving  slower than the
distance-headway allows. This is a direct effect of the
vehicle-vehicle interactions  \cite{neubert} and should therefore be
reproduced by any realistic traffic-model.

\section{Simple stochastic CA models}

Throughout this article we investigate microscopic traffic models
that are discrete in space and time. The discreteness of the model
has the advantage of allowing direct and very efficient computer 
simulations, and in particular without any further
discretization errors. The discreteness of the model, however,
also leads to some difficulties, in particular when describing
congested traffic. E.g., in congested traffic a continuous range of
typical velocities exists that depend strongly on the density.
This velocity interval is mapped on a discrete set of velocity
variables. So even for an optimal reproduction of the traffic state
an upper limit for the accuracy of the model exists.
Therefore, one has to find a compromise between the degree
of realism and the level of complexity by choosing an appropriate
discretization of the velocity.

Moreover, the temporal discretization introduces a characteristic time-scale.
This time-scale can be understood, if a parallel update is applied,
as the effective reaction-time of the drivers, which is included explicitly
in car-following models. 
Furthermore, the temporal discretization becomes obvious as peaks in the
measurement of the time-headways $t_h$. The finer the discretization
the less pronounced the peaks. 
In order to increase the resolution the time-headways in
the simulations are calculated via the relation $t_h =
\frac{d}{v}$ with the velocity $v$ of the vehicle and the
distance-headway $d$ to the preceding 
vehicle. Nevertheless, the minimal resolution is restricted 
by the discretization that determines the minimal $t_h$ difference in
free flow  $\frac{l}{\vmax}$ with the length $l$
and the maximum velocity $\vmax$ of a vehicle. In
order to facilitate a comparison with the empirical time-headway
distribution the distributions are normalized via $\sum P(t_h) \cdot
\Delta t_h = 1$.

Below we discuss a number of traffic models in detail and with
respect to their agreement with the empirical findings of our
test scenario. Beyond that we demand that each model reproduces
some basic phenomena, like spontaneous jam formation, and
fulfills minimal conditions as, e.g., being free of collisions. These
conditions are generally understood as fulfilled, if the
opposite is not explicitly stated. In particular, deterministic models
(e.g.~\cite{Fukui96,ultra}) are not a subject of this study.
They can not reproduce the spontaneous formation of jams \cite{SSNI} 
which are the result of an inherent stochasticity of traffic flow rather 
than a consequence of perturbations.

Our simulations are performed on a periodic single-lane system. This
simple structure of the system is in sharp contrast with realistic
highway networks. It is nevertheless justified, because it has been
shown for a large class of models that different boundary conditions
{\em select} different steady states rather than changing their microscopic
structure \cite{schuetz}. Therefore the boundary conditions 
are of great importance if one tries to reproduce 
the spatio-temporal structure on a macroscopic level. 
However, in comparison with local measurements an appropriate 
traffic model should be able to reproduce the empirical  
results also if periodic boundary conditions are applied.
Furthermore the restriction to a single lane is
of minor importance for the empirical test scenario which has been 
discussed in the previous section. 
In the simulations system sizes of $L\geq 10000$ cells
have been used which is sufficient to reduce finite-size effects. 
Typical runs used 50000 time steps to reach the stationary state
and measurements.

We also want to emphasize that for each model all simulations
have been performed with a  {\em single} set of parameters.
Some of the model parameters can be directly related to a given empirical 
quantity. In this case we have chosen the value
that leads to an optimal agreement with the related observable
to avoid ranking the importance of the empirical findings.
For a particular application of the model, however, the reproduction
of a certain quantity might be of special interest and therefore a
calibration of the model different from ours might be more appropriate.


\subsection{The CA model of Nagel and Schreckenberg}
\label{naschsectionl1}

The model introduced by Nagel and Schreckenberg~\cite{Nagel93} (hereafter 
cited as ``NaSch model'') is the prototype of microscopic models that
we discuss. The important role of this model is mainly
due to its  simplicity which allows  for very fast  implementations.
In fact the NaSch model  is a minimal  model in the sense that every 
further simplification leads to a loss of realism. We will also use
it as a reference for other models that will be introduced by
giving the relation to the NaSch model. 

The NaSch  model is a  discrete model  for traffic flow.   The road is
divided  into cells  that can be either  empty  or occupied by car $n$
with  a velocity $v_n=0,1,...,\vmax$.  Cars  move from the left to the
right on   a lane with periodic  boundary   conditions and  the system
update is performed in parallel. 

For completeness we repeat the definition of  the model that is
given by the four following rules ($t<t_1<t_2<t+1$):
\begin{enumerate}
\item Acceleration: $v_n(t_1) = \min\{v_n(t)+1,\vmax\}$
\item Deceleration: $v_n(t_2) = \min\{v_n(t_1),d_n(t)\}$
\item Randomization: $v_n(t+1) = \max\{v_n(t_2)-1,0\}$ with probability 
$p_{\rm{dec}}$\ \ (otherwise $v_n(t+1) = v_n(t_2)$)
\item Motion: $x_n(t+1) = x_n(t)+v_n(t+1)$
\end{enumerate}
with the velocity $v_n$, the maximum velocity $\vmax$ and the position
$x_n$ of car $n$. $d_n(t)$ specifies the number of empty  cells in front of 
car $n$ at time $t$. 

For a  given discretization the  model can be  tuned simply by varying
the two  parameters $\vmax$ and  $p_{\rm{dec}}$. The value  of
$\vmax$ mainly 
affects the slope  of the fundamental diagram  in the free flow regime
while  the behavior  in the   congested  regime is controlled  by  the
braking noise $p_{\rm{dec}}$. Each time-step $\Delta t$ corresponds 
to 1.2~s in reality in order to reproduce the empirical jam velocity at a
given cell length of $7.5$ m. The length of a cell 
corresponds to the average space occupied by a vehicle in a jam, i.e., 
its length and the distance to the next vehicle ahead. This choice 
is in accordance with measurements at German highways on the 
left and middle lane, where the density of trucks is low~\cite{kerner96}.
Due to the parallel update an implicit reaction time is introduced
which has to be considered when choosing the unit of time. This time is  
not the reaction time of the driver (that would be much shorter) but 
the time between the stimulus and the actual reaction of the vehicle.
The value we have chosen allows to reproduce the typical upstream 
velocity of a jam.

\begin{figure}[hbt]
\begin{center}
\vspace{0.3cm}
\includegraphics[width=0.9\linewidth]{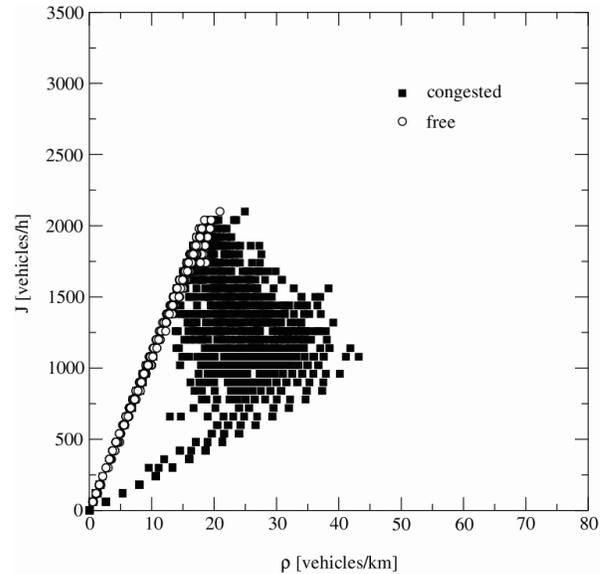}
\vspace{0.5cm}
\caption{Local fundamental  diagram   of  the NaSch  model  for
$\vmax = 112~\rm{km/h} = 5~\rm{cells}/\Delta t$, $\Delta t =
1.2$ s and $p_{\rm{dec}} = 0.16$. A cell has a length of $7.5$ m.} 
\label{nasch_fd}
\end{center}
\end{figure}

We tune the two free parameters of the model by adjusting the slope in
the free flow regime and the maximum of the fundamental diagram. 
Fig.~\ref{nasch_fd} shows the resulting fundamental diagram using 
$\vmax = 112~\rm{km/h} = 5~\rm{cells/timestep}$ and
$p_{\rm{dec}} = 0.16$  
which has to be compared with the empirical results. 

By tuning the parameters
we could reproduce quite well the free flow branch of the fundamental 
diagram: Both, the slope as well as the maximum is in agreement with 
the empirical findings. For congested traffic, however, the model 
fails to reproduce the two distinct phases, in particular the 
characteristics of synchronized traffic are not matched. This interpretation 
of the flow data is supported by measurements 
of the cross-correlation function that is negative in the corresponding
density regime.  In the presence of wide jams the flow is proportional 
to the densities as found by empirical observation. But also for wide 
jams differences exist. In real measurements the branch extends up to 
quite large densities ($\sim 70$ veh/km), while the simulation results
are restricted  to lower densities  ($\sim 40$ veh/km). 

Next we discuss the model on a microscopic level. As mentioned above
the upstream velocity of wide jams can be tuned by choosing the
appropriate discretization $\Delta t$ of the time. We have verified
our calibration by initializing the system by a large jam and
measuring the velocity of the upstream propagation of the jam front.
As expected our result is in agreement with the empirical data.
Nevertheless the dynamics of jams in the NaSch model is in
contradiction to empirical findings since its outflow from a jam
equals the maximal possible flow.  This implies that the observed
parallel propagation of jams cannot be reproduced by the NaSch model.

\begin{figure}[hbt]
\begin{center}
\includegraphics[width=0.9\linewidth]{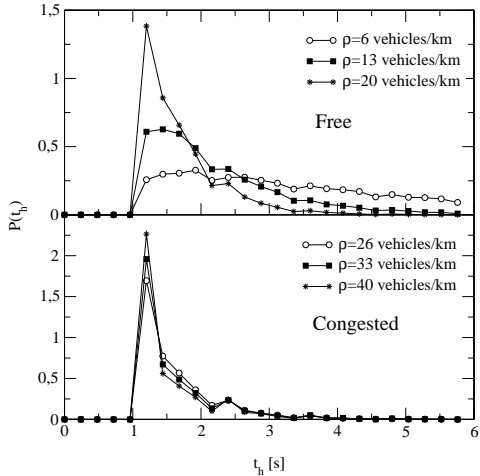}
\caption{Normalized time-headway distribution of the NaSch model in free flow 
and congested traffic for different densities.}
\label{nasch_th}
\end{center}
\end{figure}

The time-headway distributions of the NaSch model  (see also
\cite{debch}) also mismatch with empirical data
(Fig.~\ref{nasch_th}). 
Due to the discreteness of the model  
and the unique maximal velocity of the cars the distribution 
function has a peaked structure\footnote{The 
time-headway distributions have a resolution that is finer 
than the unit of time which was assigned to an update step.
This is possible because we calculate the exact passing time 
of the car from its position and velocity after executing the 
time-step. An example of a direct measurement of time-headways can 
be found in~\cite{debch}.}. 

But more important than that is the absence of time-headways shorter
than the chosen unit of time.  
This implies that we cannot reproduce the cut-off at short times 
and the upstream velocity of jams at the same time.

\begin{figure}[hbt]
\begin{center}
\includegraphics[width=0.9\linewidth]{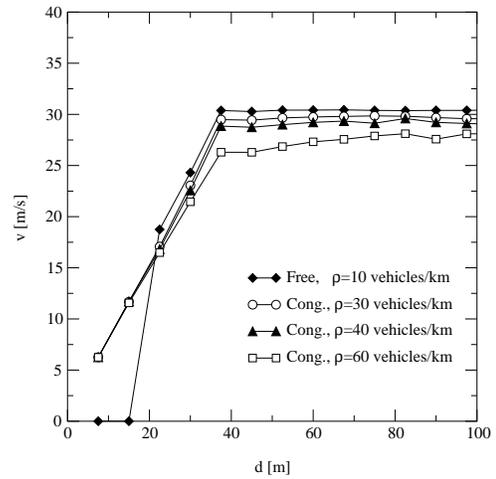}
\caption{OV function in free flow and congested traffic of the NaSch 
model for different densities.  } 
\label{nasch_ov}
\end{center}
\end{figure}

Finally, we also discuss the optimal velocity curves of the model
(Fig.~\ref{nasch_ov}). 
In congested traffic one observes only a very weak dependence 
of the ``optimal velocity'' on the density. This is due to the 
short range of interactions in the model and the strong acceleration 
of the cars.
So we neither observe a significant density dependence nor a 
sensitivity to the traffic state. This is a serious  
contradiction to the empirical findings, related to an
incomplete description of the microscopic structure of the model.


\subsection{VDR model}

A step towards a more realistic CA model of traffic flow was done
by the so-called velocity-dependent-randomization (VDR)
model~\cite{robert} that extends slightly the set of update rules 
of the NaSch model.  In this model, a velocity-dependent randomization
$p_{\rm{dec}}(v)$ 
is introduced that is calculated before application of  step 1 of the
NaSch  model. As simplest  version, a  different
$p_{\rm{dec}}$ for cars with $v=0$ was studied: 
\begin{equation}
 p_{\rm{dec}}(v) =  \left\{ \begin{array}{cc}  
     p_{0} & \ \ \text{for\ } v =  0\\
     p     & \ \ \text{for\ } v >  0 
                    \end{array} \right. 
\end{equation}
with $p_{0} > p$ (slow-to-start rule).

The additional rule of the VDR model has been introduced in order 
to reproduce hysteresis effects. This is indeed possible, because 
the new parameter $p_0$ allows to tune the velocity and outflow 
of wide jams separately. 
As a side effect it is now possible to reproduce the observed short
time-headways by keeping the  
unit of time small {\it and} the empirical observed 
downstream velocity of jams. The parameters of the model were chosen 
in the following way: The unit of time was adjusted in order 
to match the position of the maximum of the time-headway 
distribution. Then we have chosen the parameter $p_0$ such that 
we could reproduce the measurements of the upstream velocity 
of a jam. Finally the values of $\vmax$ and $p$ ensure 
a good agreement in the free flow branch. 
 The behavior found in the VDR model is typical for models
with slow-to-start rules \cite{annalen,tt}.

\begin{figure}[hbt]
\begin{center}
\vspace{0.3cm}
\includegraphics[width=0.9\linewidth]{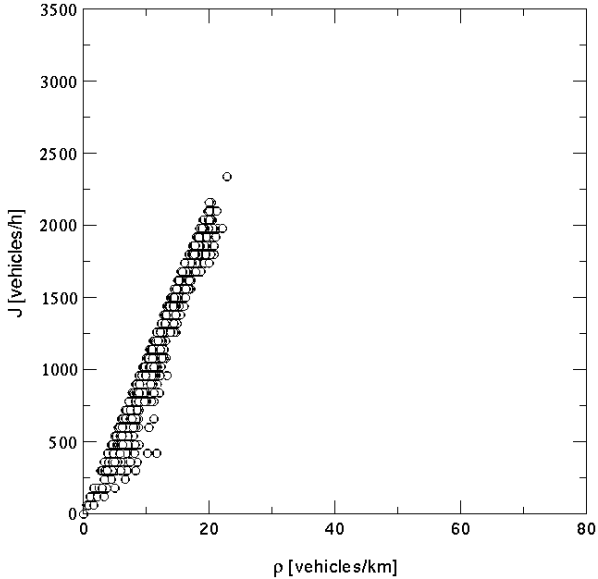}
\vspace{0.5cm}
\caption{Local fundamental diagram of the VDR model for $\vmax
  =  108~\rm{km/h}  = 3~\rm{cells}/\Delta t$, $\Delta t = 0.75~\rm{s}$,
   $p_{0}  =  0.58$ and $p  = 0.16$.} 
\label{vdr_fd_local}
\end{center}
\end{figure}

Fig.~\ref{vdr_fd_local} shows the local fundamental diagram of the VDR
model. For the parameter values obtained by the above procedure only
very weak hysteresis effects are observed.
Obviously the model fails to reproduce the empirically observed
congested phase correctly. Compared to the NaSch model the mismatch of 
the fundamental diagram in the congested regime is even more serious, 
i.e., we  cannot identify at all a density regime as synchronized traffic. 
The reason for this is a stronger separation between free flow and wide jams,
which are compact. Therefore one does not observe any flow within 
a jam if a stationary state of a periodic system is analyzed.
In case of open boundary conditions a slight broadening of the 
free-flow branch has been observed, if the detector is
located close to the exit of the highway section. This 
effect is due to the smaller length scale of jams close 
to the exit, which leads to a larger weight of accelerating 
cars. Due to the coarsening of the jam size this effect
vanishes in the bulk of the system \cite{appert,robert_neu}.

But in any case, this way of generating synchronized states by 
the boundary conditions does not agree with the empirical 
situation, because one cannot reproduce the large spatial 
and temporal extension of the synchronized state. The missing 
synchronized traffic phase leads to quite large positive values 
of  the cross-correlation $cc(J,\rho)$ of the density and the flow.

\begin{figure}[hbt]
\begin{center}
\includegraphics[width=0.9\linewidth]{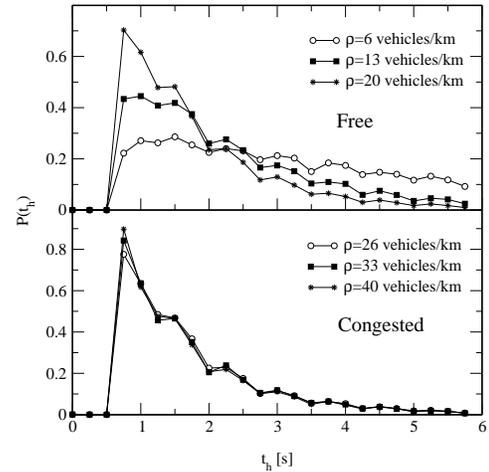}
\caption{Normalized time-headway distribution of the VDR model in free flow 
and congested traffic for different densities. } 
\label{vdr_th}
\end{center}
\end{figure}

\begin{figure}[hbt]
\begin{center}
\includegraphics[width=0.9\linewidth]{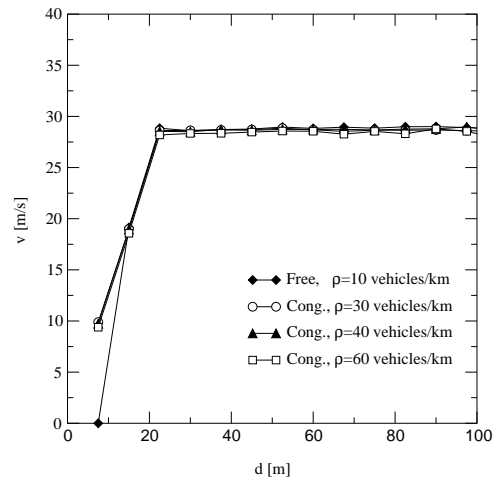}
\caption{OV function of the VDR model in free flow and congested
traffic for different densities.} 
\label{vdr_ov}
\end{center}
\end{figure}

The time-headway distribution of the VDR model differs in two 
points from the empirical observations (Fig.~\ref{vdr_th}). (i) The
unit of time is a sharp cut-off, 
i.e., the short  
time characteristics of the time-headway distribution is not 
in agreement with the empirical findings. (ii) We do not 
observe a density dependence of the maximum in congested 
traffic. Similar results are obtained for the OV functions, 
that do not depend on the density or the traffic state (Fig.~\ref{vdr_ov}). 
This result is a consequence of the microscopic structure of 
high density states. At large densities compact wide jams and zones of 
free flow traffic coexist, separated by a narrow transition layer. Now, 
our virtual ``detector'' measures only moving cars and therefore almost 
freely moving  cars even at large densities.

The major achievement of the VDR model is the correct description of 
the dynamics of wide jams  which is similar to the so-called
local cluster effect \cite{localcluster} found in hydrodynamical models.
The outflow from a jam is lower than the
maximal flow, and therefore jams do not emerge in the outflow
region. This effect leads to the increased stability of jams, 
including the empirically observed parallel upstream motion of two 
jams. 

The analysis of the VDR model showed even more clearly the effect of a 
missing synchronized traffic phase. While in the NaSch model 
the density can be chosen such that a scattered structure in the 
fundamental diagram appears, we obtain rather pure free flow 
states and wide jams for the VDR model. Contrary the VDR model gives 
a much better description of the dynamics of jams.  
In contrast to the NaSch model, the VDR model is able to reproduce 
e.g.\ the parallel motion of coexisting jams \cite{appert,robert_neu}. 
Although this phenomenon is rarely observed it should be 
reproduced by a realistic traffic model, because it is a sensitive 
for the correct description of the motion of jams. In case 
of the NaSch model this pattern is not observed, because 
new jams can form in the downstream direction  of a jam.


\subsection{The time-oriented CA model}

Based on the  CA model of  Nagel  and Schreckenberg, Brilon  {\it  et
al.}~\cite{brilon} proposed a time-oriented CA model (hereafter cited
as TOCA) that increases the interaction horizon of the NaSch model (where
cars interact only for  $d \le v$) and therefore
changes the car-following behavior. 

Compared to the NaSch model the acceleration step is modified, i.e., 
a  car accelerates only if its temporal headway $t_{h} =  d(t)/v(t)$  
is larger than some safe time-headway $t_{s}$. But even for 
sufficiently large headways the acceleration of a vehicle is not
deterministic,  
but is applied with probability  $p_{\rm{ac}}$. As a second modification 
also the randomization step is modified, i.e., it is performed only 
for cars moving with short time-headways $(t_{h} <  t_{s})$. The 
limited interaction radius of this third step leads, for a given 
value of $p_{\rm{dec}}$, to a reduction of the spontaneous jam formation. 

The update rules then read as follows $(t<t_1<t_2<t+1)$: 
\begin{enumerate}
\item   if $(t_{h} > t_{s})$   then\\  $v_n(t_1) =   \min\{v_n(t)+1,
\vmax\}$ with   probability $p_{\rm{ac}}$ 
\item $v_n(t_2) = \min\{v_n(t_1), d_n(t)\}$ 
\item if $(t_{h} <  t_{s})$  then\\ $v_n(t+1) =  \max\{v_n(t_2)-1, 0\}$  
with probability       $p_{\rm{dec}}$ 
\item $x_n(t+1) = x_n(t) + v_n(t+1)$
\end{enumerate}
with  $t_{s} = 1.2$, $p_{\rm{ac}} =  0.9$ and $p_{\rm{dec}} =
0.9$~\cite{brilon}. 
For the comparison with the NaSch and VDR model we use $v_{\rm{max}} = 4$. 

With this choice  of $t_{s}$ the   update rules can be  simplified for
$v_{\rm{max}} \le 4$, because of the discrete nature of the model:

\begin{enumerate}
\item $v_n(t_1) = \min\{v_n(t)+1, v_{\rm{max}}\}$ with probability 
$p_{\rm{ac}}$ 
\item $v_n(t_2) = \min\{v_n(t_1), d_n(t)\}$ 
\item if $(v_n(t+1) \le d_n(t))$\\ $v_n(t+1) = \max\{d_n(t)-1, 0\}$ with
probability $p_{\rm{dec}}$  
\item $x_n(t+1) = x_n(t) + v_n(t+1)$
\end{enumerate}

\begin{figure}[hbt]
\begin{center}
\vspace{0.3cm}
\includegraphics[width=0.9\linewidth]{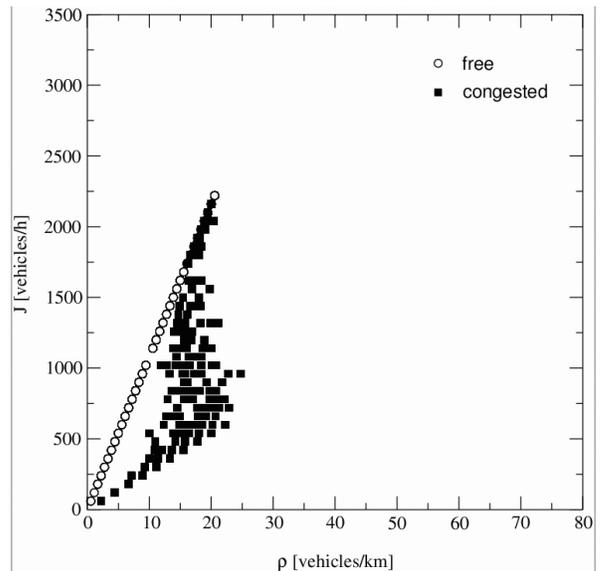}
\caption{Fundamental diagram of the TOCA model. As discretization we
used a cell length of $7.5$ m and a time-step corresponding to 
$\Delta t=1$ s in reality. The parameters of the model are chosen
as $t_{s} = 1.2$, $p_{\rm{ac}} = p_{\rm{dec}} = 0.9$ and 
$ v_{\rm{max}} = 4~\rm{cells} / \Delta t = 108$ km/h. }
\vspace{0.5cm}
\label{toca_fundi_local}
\end{center}
\end{figure}

As expected for this parameterization of the model we obtain results 
for the fundamental diagram that are similar to the NaSch model
(Fig.~\ref{toca_fundi_local}).  

The absence of spontaneous velocity fluctuations at low densities,
however, implies that $p_{\text{dec}}$ has to be chosen quite large in 
order to obtain realistic values of the maximal flow. At the 
same time large values of the braking probability lead to the 
formation of jams at low densities, such that it is difficult to 
obtain density fluctuations with amplitudes comparable to the 
empirically observed values.

\begin{figure}[hbt]
\begin{center}
\includegraphics[width=0.9\linewidth]{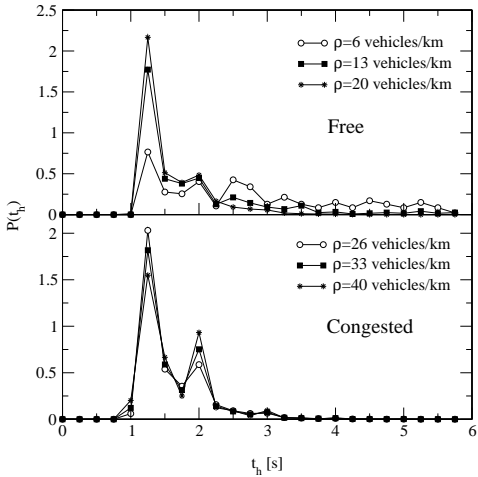}
\caption{Normalized time-headway distribution for the TOCA model in
free flow and congested traffic for different densities.} 
\label{toca_th}
\end{center}
\end{figure}

The time-headway distributions of the TOCA model, however, 
differ significantly from the results of the NaSch model
(Fig.~\ref{toca_th}). For free flow traffic the position of the maximum 
is different from the minimal time-headway for the chosen set of parameters. 
The maximum coincides with $t_h$ while the minimal time-headway is 
determined by the unit of time. 
For congested traffic the distribution has two maxima, one corresponding 
to the typical time-headway in free flow traffic and the other 
corresponding to the typical temporal distance in the outflow region 
of a jam.

\begin{figure}[hbt]
\begin{center}
\includegraphics[width=0.9\linewidth]{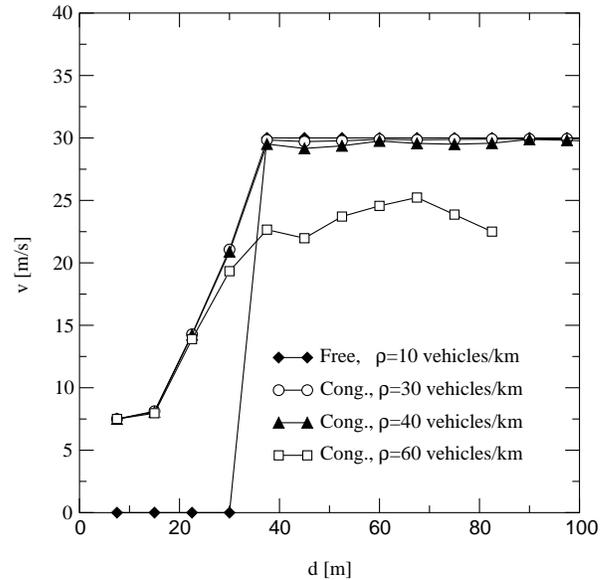}
\caption{OV functions of the TOCA model in free flow and congested
traffic for different densities.} 
\label{toca_ov}
\end{center}
\end{figure}

The OV functions of the TOCA and the NaSch model differ in two respects
(Fig.~\ref{toca_ov}). 
$(i)$ Due to the fact that the randomization step is applied for 
a finite range of the interactions, all cars move deterministically 
with $v_{\rm{max}}$ at low densities and therefore spatial headways 
smaller than $\vmax$ cells are completely avoided. This result 
is at least partly a consequence of our simulation setup, i.e., 
choosing exactly the same maximal velocity for every car. 
$(ii)$ The second difference is found in the density dependence of 
the OV-function for congested traffic.
Because of the retarded acceleration in step 1 and the deceleration of
vehicles with $v \le d$, at very large densities the system contains
only one large jam with a width comparable to the system size. 
As a consequence, the mean velocity at a given distance is reduced
considerably compared to free flow. The transition to a
completely jammed system occurs at densities of about $66$ veh/km and
leads to the abrupt change of the OV-curve.

The main difference between the NaSch model and the TOCA approach is  
the structure of jams. Due to the restricted application of 
the randomization step, $p_{\rm{dec}}$ must be quite large in 
order to obtain reasonable results for the fundamental diagram. 
Such a choice of  $p_{\rm{dec}}$, however, reduces significantly the 
density of jams. This implies that, although the typical time-headway in 
the outflow region of a jam has the correct value,
the downstream  velocity of jams is too large.

Our analysis revealed several shortcomings of the TOCA model. However,
we believe that the TOCA model is an interesting advancement of the
NaSch model if a finer spatial discretization is applied.  We
will illustrate this for the example of the density of a wide jam: The
inverse density of wide jams was used in order to fix the size of a
cell.  This choice is correct, as the jams in the model are basically
compact, which is not true in case of the TOCA model. In this case
more accurate results could be obtained if each cell would be divided
into three cells. Using this finer discretization cars occupy
two cells which would finally lead to a quite realistic dynamics of
jams. A more elaborate discussion of the discretization effects can be
found in appendix~\ref{a}.


\section{CA models with modified distance rules}
\label{sec_moddist}

\subsection{The model of Emmerich \& Rank}

The CA model introduced by Emmerich~and~Rank~\cite{emmerich} 
(ER-model) is another variant 
of the NaSch model with an enhanced interaction radius.
Precisely speaking the braking rule of the NaSch model is replaced
by applying a velocity dependent safety rule that is implemented 
via a gap-velocity matrix $M$. The entries $M_{ij}$ of $M$ 
denote the allowed velocities for a  car with gap $i$ and 
velocity $j$. Replacing the braking rule $M_{ij} \leq j$ holds because 
otherwise the car would accelerate. For the NaSch model the elements
of the gap-velocity  matrix $M^{(\rm{NaSch})}$ simply read 
$M_{ij}= \min\{i,j\}$.  

Emmerich and Rank tried to improve the NaSch model by introducing 
a larger interaction horizon, i.e. by an earlier adaption of the 
speed.  This partly avoids the unrealistic effect, that drivers 
stop from a high speed within one time step. Compared to the Nasch model 
their choice of the matrix $M$ only modifies the distance rule 
for cars moving with velocity $\vmax$: If $4\leq d \leq 9$
the car $n$ has to slow down to velocity $4$. For all other combinations
of $d$ and $v$ the NaSch distance rule is left unchanged.

As a second modification of the NaSch model a different update scheme 
is applied.
The ER model uses an unusual variant of the ordered sequential 
update, i.e., all 
rules, including the movement of the vehicles, are directly applied
for the chosen car. 
A unit of time corresponds to one update of all cars. Ordered 
sequential updates use normally a fixed sequence of cars or lattice 
sites. This has the disadvantage that some observables, e.g., the 
typical headway, may depend on the position of the detection device, 
even for periodic systems. In order to reduce this effect the 
car with the largest gap is chosen first and than the update
propagates against the driving direction~\cite{emmerich}. 

\begin{figure}[hbt]
\begin{center}
\vspace{0.3cm}
\includegraphics[width=0.9\linewidth]{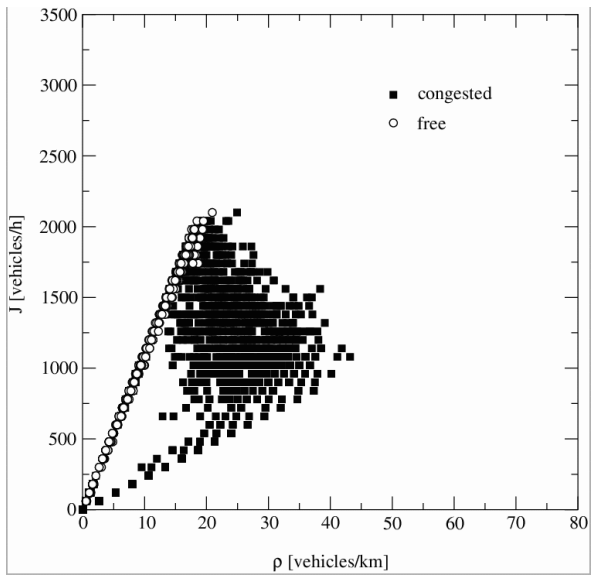}
\vspace{0.5cm}
\caption{Fundamental diagram of the ER-model. As suggested in 
the original work we have chosen $7.5$ m as the length 
of a cell, $\Delta t = 1$ s, $p_{\rm{dec}} = 0.3$ and $v_{\rm{max}} =
5 = 135$ km/h.}
\label{emmerich_fd}
\end{center}
\end{figure}

\begin{figure}[hbt]
\begin{center}
\includegraphics[width=0.9\linewidth]{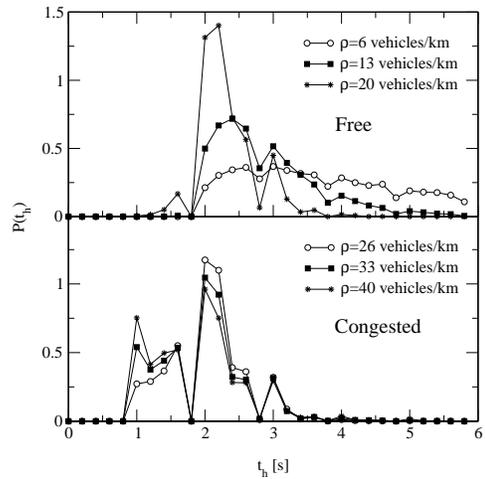}
\caption{Normalized time-headway distribution for the ER model in free
flow and congested traffic for different densities.}
\label{er_th}
\end{center}
\end{figure}

As a consequence of the ordered sequential update scheme, the gaps are
used very efficiently and very large flows can be  achieved~\cite{Rajewsky}. 
(Now, it is allowed that two cars  are driving with  $v_{\rm{max}}$ and  
$d = 0$,  so that   flows $J > 1~\rm{veh}/{\Delta t}$  are  possible). 
Therefore  large  deceleration
probabilities are necessary to decrease the  overall flow to realistic
values.   Nevertheless, due  to the    sequential  update scheme,  the
spontaneous jam formation is reduced considerably. The application of 
a sequential update is crucial. If it is replaced, e.g., 
by a parallel update, one may observe an unrealistic form, 
i.e., a non-monotonous behavior, of the fundamental diagram 
at low densities \cite{review}. 

Due to the special choice of $M^{(\rm{ER})}$, the velocity of cars
with $d \le 9$ is restricted to $v \le 4$.  This means, that a generic
speed limit with $v_{\rm{max}} = 4$ is applied for all densities $\rho
\ge 1/11 \approx 12$ veh/km, where the mean distance between the cars
is smaller than $10$ cells.  Therefore, the free flow branch of the
fundamental diagram in Fig.~\ref{emmerich_fd} has in contrast to the
empirical data two different slopes, one corresponding to
$v_{\rm{max}} = 5~\rm{cells}/\Delta t$ if $\rho < 15$ veh/km and the
other to $v_{\rm{max}} = 5~\rm{cells}/\Delta t$ at larger densities.

For the present choice of $M$ we recover basically the distance 
rule of the NaSch model with $v_{\rm{max}} = 4~\rm{cells}/\Delta t$, 
because the speed limit applies only for larger distances.  Therefore 
the structure of the congested part of the fundamental diagram 
is quite similar to the NaSch model.
However, important differences concerning the microscopic structure of 
the traffic state exist, mainly due to the modified update scheme. 
The ordered sequential update allows motion at high speeds and small 
distances.
This could in principle (for small $p_{\rm{dec}}$) 
lead to very short time-headways. For the chosen value of
$p_{\rm{dec}}$, however, 
the typical time-headways are quite large in the free flow regime and 
do not match the empirical findings. Nevertheless the ordered 
sequential update changes qualitatively the form of the time-headway 
distribution, i.e., the position of the maximum and the short time 
cut-off are different, as empirically observed (Fig.~\ref{er_th}). 

\begin{figure}[hbt]
\begin{center}
\includegraphics[width=0.9\linewidth]{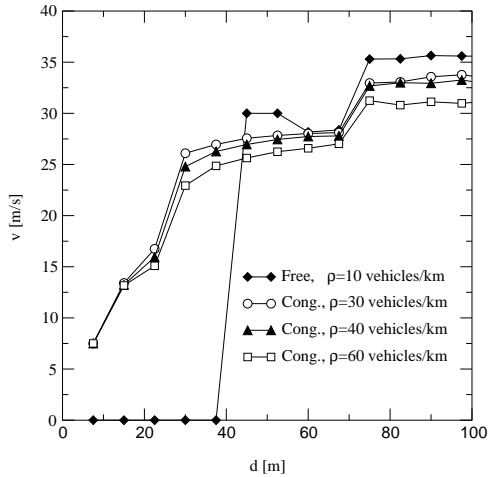}
\caption{OV function of the ER model in free flow and congested
traffic for different densities.}
\label{er_ov}
\end{center}
\end{figure}

The OV-function of the ER-model differs strongly from the empirical 
findings (Fig.~\ref{er_ov}). For this quantity the modified distance
rule is of great  
importance. In the congested regime, we observe plateaus of 
almost constant average velocities $v<v_{\rm{max}}$. The density dependence
of the OV function is, as for the NaSch model, very weak.
In free flow traffic small headways simply have not been observed, in
contradiction  to the empirical results. 

The most important weakness of the ER-model is its description of the 
jam dynamics. First of all for small values of $p_{\rm{dec}}$ the possibility 
of downstream moving jams exist, which contradicts all empirical 
studies. But even for the large value of $p_{\rm{dec}}$ we applied,
jams are not  
stable, i.e., often branch into a number of small jams. Therefore 
it is impossible to reproduce the empirically observed parallel 
moving jams with the ER-model. 

In  summary, the gap-velocity matrix  allows a more
detailed modeling of the interaction horizon. But keeping the parallel 
update scheme, unrealistic behavior at low densities is observed. 
Using a special variant of
the sequential update leads to a very unrealistic structure of 
the microscopic traffic states.


\subsection{A discrete optimal velocity model}

Helbing and Schreckenberg (HS)~\cite{HeSch} have introduced a cellular
automata (CA) model for the description of highway traffic based
on the discretization of the optimal-velocity (OV) model of Bando et al.\
\cite{bando}. The model was introduced in order to provide 
an alternative mechanism of jam formation. In certain density 
regimes the HS model is very sensitive to external perturbations 
due to its intrinsic nonlinearity. So in contrast to the previous 
approaches the pattern formation is of chaotic rather than of
stochastic nature, although the definition of the model includes 
a stochastic part as well.

The deterministic part of the velocity update is done by assigning the 
following velocity to the cars:
\begin{equation}
 v_n(t+1/2) = v_n(t) + \Big\lfloor \lambda [ V_{\rm{opt}}(d_n) - v_n(t) ]
\Big\rfloor
\label{accel}
\end{equation}
where  $V_{\rm{opt}}(d)$ denotes the ``optimal'' velocity of  
car $n$ for a given distance $d_n$ to the vehicle ahead, $v_n(t)$ the 
discrete velocity at time $t$ and $\lfloor \dots \rfloor$ the 
floor function. The constant  $\lambda$ is a free parameter of 
the model\footnote{In contrast to the original work we consider here 
only the case of one type of cars.
Furthermore we denote the OV function by $V_{\rm{opt}}$
so that it can be better distinguished from the velocities of the
cars.}.
The acceleration step is the naive discretization of the acceleration step
of the space and time continuous OV model. In the continuous version 
of the OV model the parameter $\lambda$ determines the timescale of
the acceleration. However, for time-discrete models it is well known
that a simple rescaling of time is not possible. Therefore the
meaning of the parameter $\lambda$ remains unclear.

The deterministic update is followed by a randomization step as known 
from the NaSch model, i.e. the velocity of a car with $v_n(t+1/2)>0$ is 
reduced with probability by one unit.

Although the definition of the model seems to be quite similar to the models 
discussed in the previous sections, important differences 
exist. In all other models discussed so far acceleration 
is limited to one velocity unit per time-step while breaking 
from $\vmax$ to zero velocity is possible.
This is not true for the HS model where a standing car may accelerate 
towards $\lfloor \lambda V_{\rm{opt}}(\infty) \rfloor>1$ in a single 
time-step.
On the other hand, in particular for small values of $\lambda$, the 
 braking capacity of cars is reduced. 
A reduced braking capacity, however, may lead to accidents (see the 
discussion in~\cite{krauss,HT,Sasoh}), 
a certainly unwanted feature of a traffic model. It also implies
that the model is not defined completely by the dynamics. This becomes
a problem especially in simulations. Here further rules are necessary
to determine how to deal with accidents.

\begin{figure}[hbt]
\begin{center}
\vspace{0.3cm}
\includegraphics[width=0.9\linewidth]{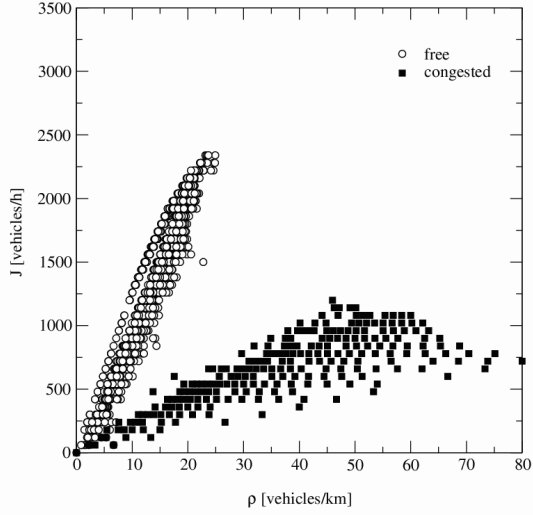}
\vspace{0.5cm}
\caption{Fundamental diagram of the HS-model. As suggested in 
the original work we have chosen $2.5$ m as the length 
of a cell, a vehicle has a length of $2$ cells, $\Delta t = 1$ s,
$p_{\rm{dec}} = 0.001$, $v_{\rm{max}} = 15 = 135$ km/h and 
$\lambda = 1/1.3$.}
\label{hs_fd}
\end{center}
\end{figure}

We will discuss the possibility of accidents in some more 
detail in appendix~\ref{b}. This discussion concentrates
on a criterion which ensures that for {\it any possible initial 
condition} no accident occurs.
 
In~\cite{HeSch}  for comparison with empirical data 
the following OV-function is suggested: 
\medskip
\begin{center}
\begin{tabular}{c|c||c|c}
$ d $ $[\Delta x]$ & $OV(d)$ $[\Delta x/\Delta t]$ & $d$ $[\Delta x]$ & 
$OV(d) $$[\Delta x/\Delta t]$ \\
\hline
0, 1 &  0 & 11     & 8  \\ 
2,3  &  1 & 12     & 9 \\
4,5  &  2 & 13     & 10 \\
6    &  3 & 14,15  & 11 \\
7    &  4 & 16--18  & 12 \\
8    &  5 & 19 -- 23  & 13 \\
9    &  6 & 24 -- 36  & 14 \\
10   &  7 & $\geq 37$  & 15 \\
\end{tabular}
\end{center}
\medskip

The length of a cell is set to $\Delta x = 2.5$ m,
$\Delta t = 1$  is chosen as the unit of time, $\lambda = \frac{1}{1.3}$ 
and we used the randomization probability  
$p_{\rm{dec}} = 0.001$ as suggested in~\cite{HeSch}.
A vehicle has a length of $l = 2~$cells corresponding to $5$~m.
For this choice of $\lambda$ and the OV function the model 
is not strictly free of collisions as our discussion in the 
appendix shows, but does at the same time not lead to 
accidents if an appropriate initial condition is chosen and
the density is not too high.
\begin{figure}[hbt]
\begin{center}
\includegraphics[width=0.9\linewidth]{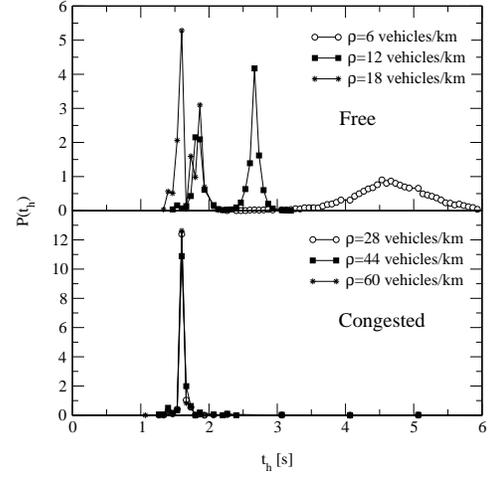}
\caption{Normalized time-headway distribution for the HS model in free
flow and congested traffic for different densities.}
\label{hs_th}
\end{center}
\end{figure}

\begin{figure}[hbt]
\begin{center}
\vspace{0.3cm}
\includegraphics[width=0.9\linewidth]{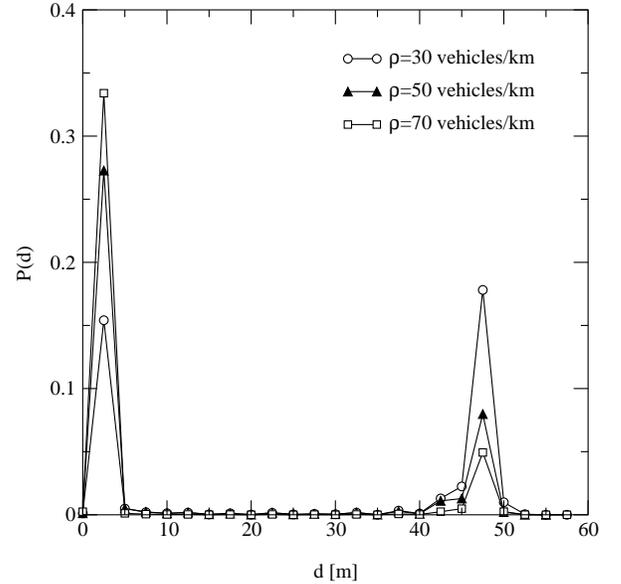}
\vspace{0.5cm}
\caption{Distance headway distributions in the congested
  regime. Obviously non-compact jams coexist with free flow 
  regimes, where the distance between two cars is rather large.}
\label{hs_dist}
\end{center}
\end{figure}

\begin{figure}[hbt]
\begin{center}
\vspace{0.3cm}
\includegraphics[width=0.9\linewidth]{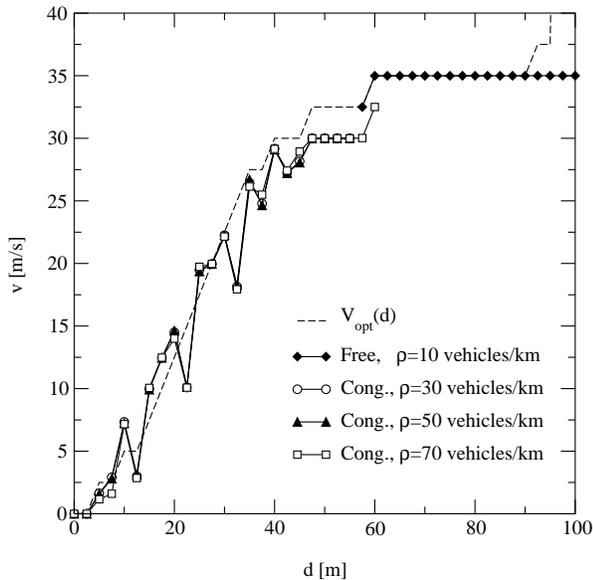}
\vspace{0.5cm}
\caption{OV function of the HS model in free flow and congested
traffic for different densities.}
\label{hs_ov}
\end{center}
\end{figure}

The optimal velocity function that governs the deterministic 
part of the vehicle dynamics, leads to speed limits in
certain density regimes $l/(d_{\rm{max}}+l) < \rho < l/(d_{\rm{min}}+l)$ 
with $V_{\rm{opt}}(d_{\rm{min}}) = V_{\rm{opt}}(d_{\rm{max}})$.
These different optimal speeds become visible in different
slopes in  the free flow branch of the fundamental diagram
(Fig.~\ref{hs_fd}).
For congested traffic two different traffic regimes 
can be identified, as empirically observed. For 
very high densities, one observes a reasonable agreement 
with the empirical data, i.e. the form of the jammed 
branch is reproduced qualitatively. This branch of the 
fundamental diagram is, however, observed only in a very 
narrow interval of global densities. 

Compared to the two other traffic states the reproduction of 
synchronized traffic is rather poor. First, one obviously observes 
a strong correlation between density and flow, which is contrast
to the empirical findings, and second, the range of densities 
which is observed in local measurements is quite narrow.

The main difficulties of the model are visible when comparing 
it with empirical results on a microscopic level. 
The simulations for the time-headway distribution show a 
strong density dependence of the maximum for the free flow states.  
This is due to the long-ranged interactions that tend to generate 
traffic states that are very  homogeneous. Therefore 
short time-headways are suppressed at low densities. The second 
problem is the quasi-deterministic character of the model. This 
implies that drivers obey the distance rule in almost any case. 
As a result the peak values of the time-headway distribution have 
extremely high weights. In congested traffic we observe a density 
independent position of the maximum of the  time-headway
distribution. The maximum carries almost the whole weight
of the distribution, in contradiction to the empirical findings. 
The reason for this can be read off from the distance headway 
distributions for different global densities (Fig.~\ref{hs_dist}).
Within a large density regime 
we observe coexistence of non-compact jams and free flow 
traffic. Therefore we can state that both high density states 
correspond to stop-and-go traffic, i.e. the model fails 
to reproduce synchronised traffic at all. 

The mismatch of the model and empirical structure of traffic 
states is also obvious for the OV-function 
(Fig.~\ref{hs_ov}). It shows almost no density dependence and 
is basically independent of the traffic 
state. The difference between the different curves is
only in a density dependent 
cut-off of the distribution, i.e. at high densities large distances 
simply do not occur.

The simulations show that HS model fails to reproduce the microscopic 
structure of the empirical observed traffic states. From our point 
of view the problems describing the empirical observation are due 
to the nature of the model. It introduces a static rule that
leads to a reasonable agreement with the empirical fundamental 
diagram. For a proper choice of $\lambda$ the vehicles take 
instantaneously a velocity close to the optimal velocity, 
i.e.\ the dynamical aspects of highway traffic are extremely simplified. 
Therefore inhomogeneous traffic states are only observed in the presence
of quenched disorder \cite{HeSch}, e.g. different types of cars, and not 
produced spontaneously.

\section{Brake light version of the NaSch model}
\label{sec_bl}

Quite recently a brake light (BL) version of the NaSch model has been
introduced ~\cite{knospe2001,knospe2002} 
in order to give a more complete description of the 
empirically observed phenomena in highway traffic. 
In contrast to the 
models we considered in the previous sections, which represent already 
well known modeling approaches,  we also discuss the basic features 
of the model that have not been presented so far. 
In the development of the model the main aim was the reproduction
of the empirical microscopic data in a robust way.


\subsection{Definition of the BL model}

The BL model combines several elements of older modeling approaches, 
e.g.,  velocity anticipation~\cite{barret,wolfgang} and a
slow-to-start rule \cite{tt,robert}. In addition,  
a dynamical long ranged interaction is included: In their 
velocity dependent interaction horizon drivers react on brakings of 
the leading 
vehicle that are indicated by an activated brake
light~\cite{bremslichtpaper}. 
The interaction, however, is limited to nearest neighbor
vehicles~\cite{HermanRothery}. 
The update rules are formulated in analogy to the VDR model. In particular 
the interactions are strictly local and a parallel update scheme 
is applied. 

In order to allow for a finer spatial discretization 
for a given length of a car, 
we include the possibility that a car may occupy more than 
a single cell. Therefore the gap between consecutive cars 
 is given by $d_n=x_{n+1}-x_n-l$ (where $l$ is  the length of the cars).
The brake light  $b_n$  can take on two states, i.e., on
(off) indicated by $b_n=1 (0)$.   In our approach the randomization
parameter $p_{\rm{dec}}$ for the  $n$th car  can take  on three
different values 
$p_0$, $p_d$ and $p_b$, depending on its current velocity $v_n(t)$ and
the status  $b_{n+1}$  of  the  brake light of   the preceding vehicle
$n+1$: 
\begin{eqnarray}
p_{\rm{dec}}  &=& p_{\rm{dec}}(v_{n}(t),b_{n+1}(t),t_h,t_s) \nonumber\\
&=& \left\{\begin{array}  {llll} 
        p_b & &{\rm  if\ }b_{n+1}=1  {\rm \ and\ } t_{h}  <  t_{s}\\
 p_0 &  &{\rm if\ 
      }v_n=0\\ p_d & &{\rm in\ all\ other\ cases}. 
\end{array} \right.
\end{eqnarray}

The two times $t_{h} = \frac{d_{n}}{v_{n}(t)}$ and $t_{s} =
\min\{v_{n}(t),h\}$, where $h$ determines the range of interaction
with the brake light, are the time $t_{h}$ needed to reach the
position of the leading vehicle which has to be compared with a
velocity-dependent (temporal) interaction horizon $t_{s}$.  $t_{s}$
introduces a cutoff that prevents drivers from reacting to the brake
light of a predecessor which is very far away.  Finally
$d_n^{(\rm{eff})}=d_n+\max\{v_{\rm{anti}}-d_{\rm{security}}, 0\}$
denotes the {\em effective} gap where $v_{\rm{anti}} =
\min\{d_{n+1},v_{n+1}\}$ is the expected velocity of the leading
vehicle in the next time-step. The effectiveness of the anticipation
is controlled by the parameter $d_{\rm{security}}$.  Accidents are
avoided only if the constraint $d_{\rm{security}}\geq 1$ is fulfilled.
The update rules then are as follows ($t<t_1<t_2<t+1$):

\begin{enumerate}
\item[0.]    Determination  of the  randomization   parameter:\\
$p_{\rm{dec}} = p_{\rm{dec}}(v_n(t),b_{n+1}(t),t_h,t_s)$\\ 
$b_n(t+1) = 0$\\

\item[1.] Acceleration:\\ if  $\bigl((b_{n+1}(t) = 0)$  and $(b_{n}(t) = 0)
\bigr)$ or $(t_{h} \ge t_{s})$ then\\
\hspace*{0.3cm} $v_n(t_1) = \min\{v_n(t)+1,v_{\rm{max}}\}$\\ 

\item[2.] Braking  rule:\\ $v_n(t_2) =  
\min\{d_n^{(\rm{eff})},v_n(t_1)\}$\\
if $(v_n(t_2) < v_{n}(t))$ then\\
\hspace*{0.3cm} $b_n(t+1) = 1$\\ 

\item[3.] Randomization, brake:\\ 
if $({\rm rand}()  < p_{\text{dec}})$ then $\{$ \\
\hspace*{0.3cm} $v_{n}(t+1)
=\max\{v_n(t_2)-1,0\}$\\ 
\hspace*{0.3cm} if $((p_{\text{dec}} = p_{b}) \textrm{ and } 
     (v_n(t+1)=v_n(t_2)-1))$ then\\
\hspace*{0.5cm} $b_{n}(t+1) = 1 \}$ \\ 

\item[4.] Car motion:\\ $x_{n}(t+1) =x_{n}(t)+v_n(t+1)$\\ 
\end{enumerate}
Here rand$()$ denotes a uniformly distributed random number from
the interval $[0,1]$.

The new velocity of the vehicles is  determined by steps $1-3$, while step
$0$  determines the dynamical parameters of the model.  Finally,  the
position  of  the car  is shifted in accordance with the calculated
velocity in step 4. 

In order to illustrate the details of the approach we now discuss the
update rules step-wise.  

\begin{enumerate}
\item[0)] The  braking parameter $p_{\rm{dec}}$ is  calculated. For a
stopped car 
the value $p_{\rm{dec}} = p_0$ is applied.  Therefore $p_0$ determines
the upstream velocity of the downstream front of a jam. 

If the brake light of the car in front is switched on and it is found
within the interaction horizon $p_{\rm{dec}} = p_b$ is chosen.  A car
perceives a brake light of the vehicle ahead within a time dependent
interaction horizon $t_{s} = \min\{v_{n}(t),h\}$ where $v_n(t)$ is the
current velocity and $h$ an integer constant.  The velocity dependence
takes into account the increased attention of the driver at large and
reduces the braking readiness at small velocities.  This reaction is
performed only with a certain probability of $p_{b}$.  In order to
obtain a finite range of interactions a cutoff at a horizon of $h$
seconds is made~\footnote{Indeed, increasing $p_{\rm{dec}}$ to $p_b$
  is the simplest possible response to the stimulus brake light. More
  sophisticated response functions like a direct reduction of the
  velocity or the gap are conceivable but lead to some problems in
  combination with anticipation.  In addition, one can think of
  different implementations of the brake noise $p_{b}$.  For example,
  we have tried more sophisticated $p_{b}$-functions, like a linear
  relationship between $p_{b}$ and the velocity, the difference
  velocity to the predecessor or the gap, but for the sake of
  simplicity in this paper we will focus on a constant $p_{b}$.}.

Finally, $p_{\rm{dec}} = p_d$ is chosen in all other cases. 

\item[1)] The velocity of the car is increased by one unit (if it
does not already  move  with  maximum  velocity).  The  car  does  not
accelerate if its own brake light or that of its predecessor is on and
the next car ahead is within the interaction horizon. 
  
\item[2)] The   velocity of  the car  is  adjusted  according  to the
effective gap. 

The brake light of a vehicle is activated only if the new velocity is
reduced compared to the preceding time-step.
Note that the application of the braking rule does not necessarily 
lead to a change of the velocity, as it can compensate a previous 
acceleration. The restriction stabilizes dense traffic flows.

\item[3)] The  velocity  of the  car is reduced   by one unit  with  a
  certain probability $p_{\rm{dec}} =
p_{\rm{dec}}(v_{n}(t),b_{n+1},t_h,t_s)$. If the car brakes due 
  to  the predecessor's brake light, its  own  brake light is switched on. 
We also stress the fact that even for distances $d_n<h$
the action of the brake light is 
restricted to brakings that are induced by the vehicle in front
(either by the braking rule or by an activated brake light) and not 
by spontaneous velocity fluctuations.

\item[4)] The position of the car is updated. 

\end{enumerate}


\subsection{Calibration of the model}

The following parameters of  the model allow to adjust the simulation 
data to the empirical findings: the maximal velocity 
$v_{\rm{max}}$, the car length $l$, the braking parameters $p_d$,  $p_b$,
$p_0$,  the cut-off $h$ of interactions, and the minimal security
gap $d_{\rm{security}}$. The parameters of the have been chosen such 
that they can easily be related to the empirical findings. As in 
the previous models a single set of parameters is used for all traffic
states.

In order to obtain realistic values of the acceleration behavior of a
vehicle, the cell length of the standard CA model is reduced to a length
$l$ of $1.5$ m. 
Since the time-step is kept fixed at a value of $1$ s this leads to a
velocity discretization of
$1.5$ m/s which   is  of the same order as the ''comfortable'' 
acceleration of
somewhere  about $1~\rm{m/s}^{2}$~\cite{ite}.
Like in the standard CA model a vehicle has a length of $7.5$ m that
corresponds to $5$ cells at the given discretization (see
appendix~\ref{a} for a discussion of the discretization effects).

Some of the parameters can be fixed as, e.g.,  in the VDR-model: 
The  maximum velocity $v_{\rm{max}}$  is determined   by the  slope  of the
free flow  branch of the  fundamental  diagram. The upstream velocity 
of a jam can be tuned by the parameter  $p_0$ and the strength of 
fluctuations that are controlled by the parameter $p_d$ determine 
the maximal flow.

The other parameters of the model are connected with an interaction 
that have not been included in the models we discussed so far. 
The parameter  $h$ describes the   horizon above which  driving is not
influenced by  the leading  vehicle. Several empirical  studies reveal
that  $h$ corresponds  to a  {\em  temporal} headway rather  than to a
spatial one.  The estimates  for $h$ vary  from $6$ s~\cite{george},
$8$ s~\cite{miller,schlums},  $9$ s~\cite{hcm}  to $11$ s~\cite{edie}.
Another estimation for $h$  can be obtained  from the analysis of  the
perception sight distance. The  perception sight distance is  based on
the first  perception of an object in   the visual field at  which the
driver  perceives  movement   (angular  velocity).    In~\cite{pfefer}
velocity-dependent perception sight distances are presented that, for
velocities up to $128$ km/h, are larger than $9$ s.  We therefore have
chosen   $h$ to  be  $6$ s  as a  lower   bound for  the time-headway.
Besides, our simulations  show,    that a good agreement with
empirical data can   only  be
obtained for  $h \ge 6$. This corresponds  to a maximum  horizon of $6
\times 20$ cells or a distance of $180$ m at velocity $v_{\rm{max}}$. 

The next parameter one has to fix is $p_{b}$. This parameter controls 
the propagation of the brake light. A braking car in front is indeed 
a strong stimulus to adjust the own speed. Therefore $p_{b}$ has typically 
a  high value. Finally, $d_{\rm{security}}$ tunes the degree of the velocity 
anticipation and has a strong influence on the cut-off of the time-headway 
distribution.

\subsection{Validation of the full model}

With this parameter set we have  calibrated the model to the empirical
data. Leaving $p_{0}$,   $h$  and $v_{\rm{max}}$   fixed, we  got  the best
agreement with the empirical data for $p_{\rm{dec}}  = 0.1$, $p_{b} = 0.94$
and $d_{\rm{security}} = 7$.

\begin{figure}[hbt]
\begin{center}
\vspace{0.5cm}
\includegraphics[width=0.9\linewidth]{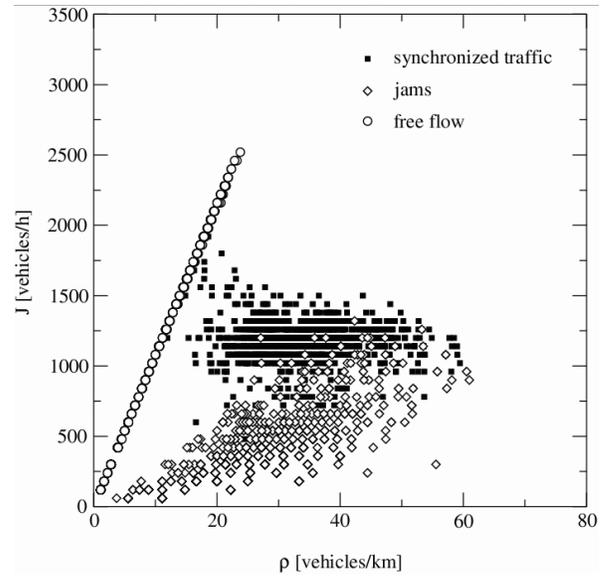}
\vspace{0.5cm}
\caption{Local fundamental diagram obtained by
  the simulation of the brake light version of the NaSch model.
  The parameters are: $p_{0} = 0.5$, $h = 6$, $v_{\rm{max}} = 20$,
$p_{\rm{dec}} 
= 0.1$, $p_{b} = 0.94$, $d_{\rm{security}} = 7$. A time-step corresponds
to $\Delta t = 1$ s, a cell has a length $l = 1.5$ m and a vehicle
covers $5$ cells.}
\label{carnaschlocal}
\end{center}
\end{figure}

As one can see in Fig.~\ref{carnaschlocal}
the  slope of  the  free flow branch and the   maximum flow
coincides  with the empirical  data indicating  that $v_{\rm{max}}$ and
$p_{\rm{dec}}$ have been   chosen properly.     

However, the   simulated densities are less distributed than in the 
empirical data set. The width of the density distribution is of
the same order as it was found for the NaSch and ER-model. The
mismatch between simulation and empirical results of the density can 
be related to discretization effects, which introduce an upper limit
for the density if simple (virtual) counting loops are used as detection
devices. 

A second lower branch appears   for  small values   of the  flow which
represents wide jams.  Because only moving cars are measured
by the inductive loop large densities cannot be calculated,
as in the empirical data of Fig.~\ref{fig:emp_fund}.

\begin{figure}[hbt]
\begin{center}
\includegraphics[width=0.9\linewidth]{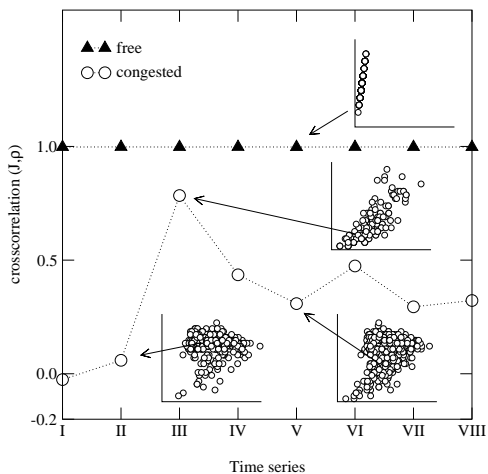}
\caption{Cross-correlation  of the flow  and the density  in free flow
and congested traffic for different densities and homogeneous
initialization.}  
\label{crosscovariance}
\end{center}
\end{figure}

The   next parameter that  can be   directly  related to an empirical
observable  quantity, namely the upstream velocity of the downstream
front of a wide jam, is  the deceleration  probability $p_{0}$.

We used the calculation of the density autocorrelation function in the
congested state of a system that was initialized  with a mega jam for
the determination     of  the  velocity  of   the   jam  front.  
One obtains an average jam velocity
of $2.36$ cells/s $(\hat{=} 12.75~\rm{km/h})$ for  $p_{0} = 0.5$. This
jam velocity is independent of the traffic condition and holds for all
densities in the congested regime.
Thus, although metastable traffic states can be achieved by the finer 
discretization (see appendix~\ref{a}) the
slow-to-start rule is necessary for the reduction  of the jam velocity
from about $20.45$ km/h  to $12.75$ km/h.   This  velocity is  also in
accordance with empirical results~\cite{kerner96}. 

In Fig.~\ref{crosscovariance} the cross-covariance $cc(J,\rho)$ of the
flow and the local measured density for different traffic states is
shown.  In the free flow regime the flow is strongly coupled to the
density indicating that the average velocity is nearly constant.  Also
for large densities, when wide jams are measured, the flow is mainly
controlled by density fluctuations. In the mean density region there
is a transition between these two regimes.  At cross-covariances in
the vicinity of zero the fundamental diagram shows a plateau.  Traffic
states with $cc(J,\rho) \approx 0$ were identified as synchronized
flow~\cite{neubert}.  In the further comparison of our simulation with
the corresponding empirical data we used these traffic states for
synchronized flow data and congested states with $cc(J,\rho) > 0.7$
for data of wide jams.  The results show that the approach leads to
realistic results for the fundamental diagram and that the model is
able to reproduce the three different traffic states.

To characterize   the   three traffic   states,  we    calculated  the
autocorrelation of the flow, the density as well as the velocity   for
different global  densities. 

In free flow, the density and the flow show the same oscillations of
the autocorrelation function, whereas the speed is not correlated in
time.

\begin{figure}[hbt]
\begin{center}
\includegraphics[width=0.9\linewidth]{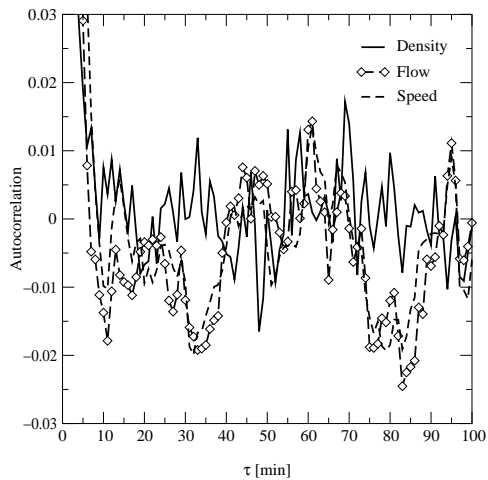}
\caption{Autocorrelation  function of  the density, the velocity
  and the flow for $\rho = 67$ veh/km with a random initialization.} 
\label{autocorrnewd5}
\end{center}
\end{figure}

In contrast to the NaSch model, the
autocorrelation function at large densities  shows a strong coupling of
the flow and the velocity. Now, the velocity of
a car not only depends on the gap but also on the density, so that the
flow  and   the   velocity   are      mainly  controlled   by      the
density (Fig.~\ref{autocorrnewd5}).

Next  we compare   the  empirical data  and  simulation  results  on a
microscopic level. 

\begin{figure}[hbt]
\begin{center}
\vspace{0.3cm}
\includegraphics[width=0.9\linewidth]{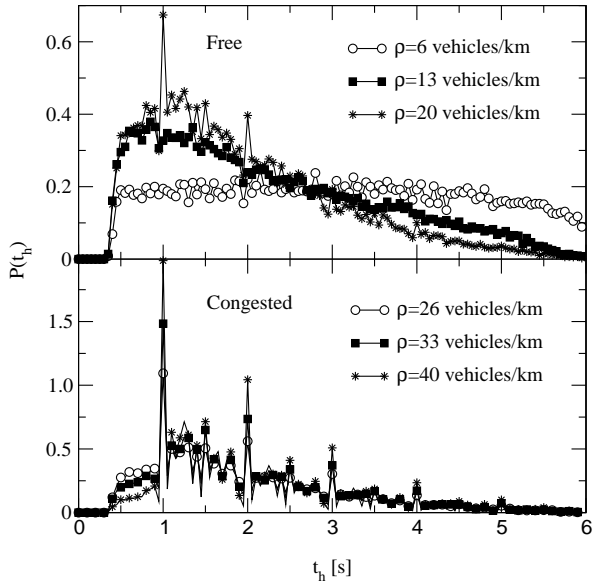}
\vspace{0.5cm}
\caption{Time-headway distribution   for  different  densities
in free flow (top) and in the synchronized state (bottom).}
\label{carnaschtime}
\end{center}
\end{figure}

In Fig.~\ref{carnaschtime}   the   simulated  time-headway
distributions  for  different density regimes are shown.   

Due to the discrete nature  of the model, large fluctuations
occur and the continuous part of the empirical distribution shows a
peaked structure at integer-numbered headways for the simulations.  In
the free flow state extremely small time-headways 
have been  found,  in accordance with the empirical results.
This is qualitatively different from the other CA models 
with parallel update scheme.

\begin{figure}[hbt]
\begin{center}
\includegraphics[width=0.9\linewidth]{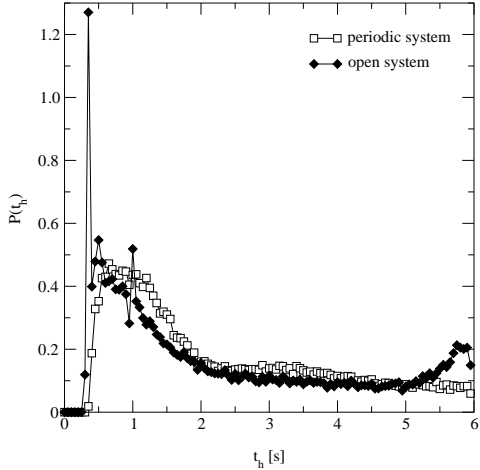}
\caption{Time-headway distribution in the free flow regime of
  a  system  with  open boundary  conditions and  different  types  of
  vehicles. The maximal  velocity of the  slow vehicles was set to
  as $v_{\rm{max}} = 108~\rm{km/h} = 20$ cells/s and of  the fast
vehicles as 
  $v_{\rm{max}}  =  135~\rm{km/h} = 25$ cells/s.  We  considered
$15\%$ of the 
  vehicles   as fast vehicles   (note   that these are vehicles  that
  disregard the speed limit).} 
\label{carnaschmixed}
\end{center}
\end{figure}

Nevertheless, for our standard simulation setup at small densities the
statistical weight of these small time-headways is significantly
underestimated. 
This apparent failure of the model is the result of the chosen 
simulation setup. If we introduce different types of cars and 
open boundary conditions, we observe a smooth time headway
distribution, which is in good agreement with the empirical data
(see Fig.~\ref{carnaschmixed}).  
Therefore we can state that properties like the width 
as well as the smoothness of the time-headway distribution 
are strongly dependent on the choice of the  simulation setup. 
In contrast, the short-time cut-off of the distribution is model
dependent. Time headways shorter than the chosen unit of time 
are in case of a parallel update only observed if anticipation effects
are included. The actual value of the cut-off for a given unit of
time is tuned by the parameter
$d_{\rm{security}}$. The results  for congested flow, however, are
not influenced by different types of vehicles. 

The  ability to anticipate  the  predecessor's behavior becomes weaker
with  increasing density so that the weight of the small
time-headways is reduced considerably in 
the  synchronized state. The maximum of the
distribution can be found in the vicinity of $1$ s in accordance with
the empirical data, the density dependence, however, cannot be
reproduced.

\begin{figure}[hbt]
\begin{center}
\vspace{0.5cm}
\includegraphics[width=0.9\linewidth]{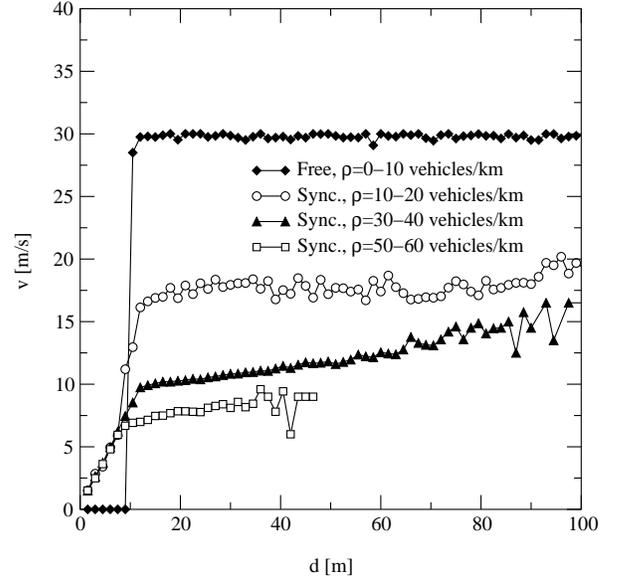}
\vspace{0.5cm}
\caption{The OV-function for different densities in the free flow and 
congested regime. }
\label{carnaschov}
\end{center}
\end{figure}

Instead, with increasing density the maximum at a time of $1$ s (in
the NaSch model the minimal time-headway is restricted to $1$ s
because of rule 2) becomes more pronounced.  This result is also due
to the discretization of the model that triggers the spatial and
temporal distance between the cars.  Because of the exponential decay
of the waiting time distribution of cars leaving a jam, the peak at a
time of $1$ s is the most probable in the time-headway distribution.

The OV curve of our model approach shows an excellent agreement with
empirical findings. For densities in the free flow regime it is obvious 
that the OV-curve (Fig.~\ref{carnaschov}) deviates from the linear 
velocity-headway curve of the NaSch model. 
Due to anticipation effects, smaller distances occur, so that driving 
with $v_{\rm{max}}$ is possible even within very small headways. 
This  strong anticipation becomes  weaker  with increasing density and
cars tend to  have smaller velocities  than the headway allows so that
the OV-curve  saturates for large  distances.  

\begin{figure}[hbt]
\begin{center}
\includegraphics[width=0.9\linewidth]{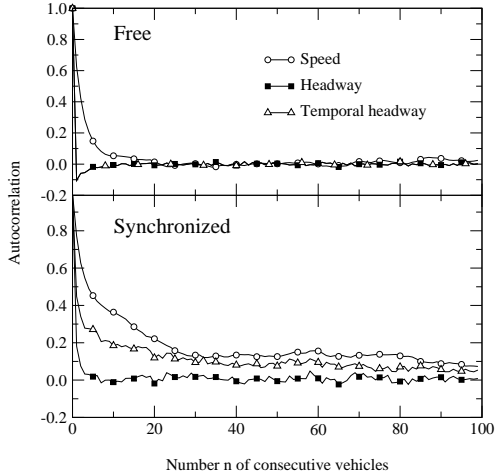}
\caption{Autocorrelation of
  the speed and of the  spatial and temporal  headway for free flowing
  vehicles (top)  and for a synchronized state  (bottom). In  order to
  obtain a  slow decay of the  speed  autocorrelation function  in the
  free flow regime the simulation was performed on  a open system with
  $20\%$  of  slow cars ($v_{\rm{max}}^{\rm{slow}}   =
15~\rm{cells/s}  = 81$ km/h,
  $v_{\rm{max}}^{\rm{fast}} = 20~\rm{cells/s} = 108$ km/h.)} 
\label{autocorrnewcar}
\end{center}
\end{figure}

The saturation of the velocity, which is characteristic for
synchronized traffic, was not observed in earlier approaches. The
value of the asymptotic velocities can be adjusted by the last free
parameter $p_b$. The OV-curve in the synchronized regime is
independent of the maximum velocity and is only determined by the
dynamical behavior of the model.

Next,  we calculated  the  autocorrelation of  the time-series of the
single-vehicle data  (Fig.~\ref{autocorrnewcar}).  Note, that the data
of the free flow state was collected in  an open system with $20\%$ of
slow cars with $\vmax =  15$ cells/s =  $81$ km/h.  In the free flow
regime the data  shows a strong coupling of  the spatial  and temporal
headway that supports the results obtained  by aggregated data
($J \propto  \Delta  t^{-1}$ and $\rho
\propto \Delta x^{-1}$).  In  contrast, the autocorrelation 
of the velocity   shows a  slow  asymptotic decay.  This supports  the
explanation  of~\cite{neubert}   that the  slow  decrease  for   small
distances is due to  small platoons of fast  cars led by one slow car.
In the synchronized  state, longer correlations  of the speed  and the
spatial headways can be observed. So, similar to the free flow regime,
in the  synchronized regime platoons of cars  appear  that are moving
with the same speed~\cite{knospe2002b}.


\section{The model of Kerner, Klenov and Wolf} 

The most recent modeling approach we include in our comparison was 
introduced by Kerner, Klenov and Wolf (KKW)~\cite{kkw}. This model 
is a fully discretized version of the space-continuous 
microscopic model introduced by Kerner and Klenov~\cite{kk}. 
It combines, as the BL model, elements of car-following theory 
with the standard distance-dependent interactions. It is defined 
by an update rule including a deterministic and a stochastic
part. The deterministic rule, which reads ($t<t_1<t+1$)
\begin{equation}
v_n (t_1)=\max\biggl\{0, \min\{v_{\rm max},  v_{\text{safe}}(t), 
v_{\text{des}}(t)\}\biggr\},
\label{next}
\end{equation}
is applied first. The three velocities appearing are the free flow or 
maximal speed of the cars $v_{\rm max}$, the safe velocity 
$v_{\text{safe}}(t)$ and finally the desired velocity $v_{\text{des}}(t)$. 
$v_{\text{safe}}(t)$ is the velocity 
which guarantees collision-free motion and is simply the gap to the 
preceeding car, $v_{\text{safe}}(t)=d_n(t)$.
It is the introduction of $v_{\text{des}}(t)$ which makes
the difference to the NaSch model. The velocity $v_{\text{des}}(t)$
is given by 
\begin{equation}
v_{\text{des}}(t)=\left\{
\begin{array}{ll}
v_{n}(t)+a         &  \quad \textrm{for } d_{n}> D (v_n(t))-l ,\\
v_{n}(t)+\Delta(t) &  \quad \textrm{for } d_{n} \leq D(v_n(t))-l. \\
\end{array} \right.
\label{next1}
\end{equation}
The calculation of $v_{\text{des}}(t)$ replaces the acceleration step 
of the NaSch model by a more complex rule.
Here $l$ is the length of the vehicles and $D(v)$ a synchronization
distance. The authors suggested a linear 
\begin{equation}
D(v)=D_{0} + kv.
\label{D1}
\end{equation}
and a quadratic form
\begin{equation}
D(v)=D_{0} + v+\beta v^{2}
\label{D2}
\end{equation}
for the velocity dependent interaction
range. Apart from the task of choosing an appropriate function, 
two model parameters are introduced in both cases. So far no
systematic analysis of traffic data exist which leads 
empirically based parameter values or the functional forms.  
This could be done, at least in principle, by an extensive analysis of 
floating-car measurements.
Moreover, the results in \cite{kkw} show that the results 
agree at least qualitatively for both functions (\ref{D1}), (\ref{D2})
which have been considered.

The interaction range has been introduced as a synchronization 
radius, i.e. $D(v)$ is the distance which separates free driving 
cars from cars which already adjust their velocity according to the
vehicle ahead. For large distances to the vehicle ahead, 
$d_{n}> D(v_n(t))-l $, the calculation 
of $v_{\text{des}}$ is equivalent to the acceleration step of the NaSch 
model. Inside the enlarged interaction radius, however, $v_{\text{des}}$ 
depends on the velocity of the leading car. 
Explicitely $\Delta(t)$ is given by 
\begin{equation}
\Delta(t)=\left\{
\begin{array}{ll}
-b &  \quad \textrm{if $v_{n}(t) > v_{n+1}(t)$} \\
0 &  \quad \textrm{if $v_{n}(t)= v_{n+1}(t)$} \\
a &  \quad \textrm{if $v_{n}(t)< v_{n+1}(t)$},
\end{array} \right.
\label{next2}
\end{equation}
This means that within the interaction radius drivers  tend to adapt
their velocity to the vehicle ahead. 

The second update rule is stochastic. It is given by 
\begin{equation}
v_{n}(t+1)=\max\bigl\{0,\min\{v_{n}(t_1)+\eta_{n}, v_{n}(t_1)+a, v_{\rm free}, 
v_{\text{max}}\}\bigr\},
\label{final}
\end{equation}

The stochasticity is included in the term $v_{n}(t_1)+\eta_{n}$, 
while the others are in order to guarantee that the new velocity is below
the speed limit, leads to no collisions and is in accordance
with the chosen acceleration capacity $a_n$ of the cars.
The stochastic variable $\eta$ can take the following values: 
\begin{equation}
\eta=\left\{
\begin{array}{ll}
-1 &  \textrm{\ \ if $r< p_{\rm b}$}, \\
1 &  \textrm{\ \ if $p_{\rm b}\leq r<p_{\rm b}+p_{\rm a}$}, \\
0 &  \textrm{\ \ otherwise}
\end{array} \right.
\label{noise}
\end{equation}
Both probabilities $p_{\rm a}$ and  $p_{\rm b}$ introduced here
are velocity dependent. One has 
 \begin{equation}
p_{{\rm b}}(v)=\left\{
\begin{array}{ll}
p_{0} & \textrm{\ \ if $v=0$} \\
p &  \textrm{\ \ if $v>0$}.
\end{array} \right.
\label{prob_b}
\end{equation}
with $p_0> p$. The stochastic braking is analogous to the 
slow-to-start rule known from the VDR model. Contrary the stochastic
acceleration is a new feature of the model which weakens the 
synchronization of speeds as it applies 
to cars which reduced or kept their velocity although safe driving 
would have allowed a larger velocity. The function $p_a(v_n)$ is
explicitly given by 
\begin{equation}
p_{\rm a}(v)=\left\{
\begin{array}{ll}
p_{\rm a 1} & \textrm{\ \ if $v <   v_{\rm p}$} \\
p_{\rm a 2} & \textrm{\ \ if $v\geq v_{\rm p}$},
\end{array} \right.
\label{prob_a}
\end{equation}
where $v_{\rm p}$,  $p_{\rm a1}$ and $p_{\rm a2}<p_{\rm a1}$ are
adjustable parameters of the model. The different probabilities 
have to be chosen such that 
$p_{a}+p_{b}\leq 1$ is fulfilled for any velocity. 
The velocity update is completed by this second stochastic 
rule and is followed by a parallel update of the positions.

For further illustration of the update rules we compare them briefly 
to the BL model. Both models include the update rules of the 
VDR model and enlarge the interaction radius of the drivers 
within a velocity dependent interaction range. The driving strategy 
within this larger interaction range is, however, different. 
While the BL model introduces an event driven interaction model, 
the KKW is more car-following like. Another important difference 
is that the velocity anticipation is not included in the approach 
of \cite{kkw},  although such an extension is possible \cite{kk2}.

\begin{figure}
\begin{center}
\vspace{0.3cm}
\includegraphics[width=0.9\linewidth]{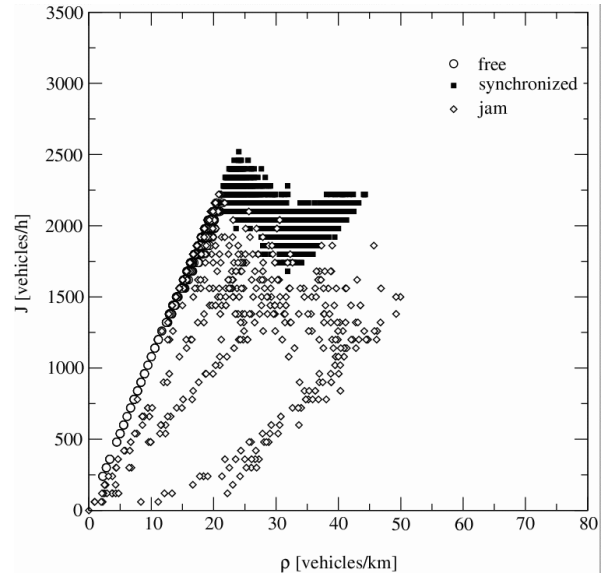}
\vspace{0.5cm}
\caption{Local fundamental diagram of the KKW model for the 
following set of parameters: The length of a cell is set to
 $0.5~\rm{m}$. Each car occupies $l=15$ cells. The maximal 
velocity is given by $\vmax = 108~\rm{km/h} = 60~\rm{cells}/\Delta t$, 
where $\Delta t = 1~\rm{s}$. Also the other model parameters 
are set to the values suggested in~[21]: 
$a = b = 1$, $D_0 = 60$, $k=2.55$. The parameters determining 
the stochastic part of the model take the values: 
$p  = 0.04$, $p_{0} = 0.425$, $p_{\text{a}1} = 0.2$, $p_{\text{a}2} = 0.052$ 
and $v_p =28$. }
\label{fig:kkwlocalfund}
\end{center}
\end{figure}

Fig.~\ref{fig:kkwlocalfund} shows the fundamental diagram 
of the KKW model, obtained by local measurements flux and 
density in a periodic 
system. Compared to the other models we analyzed one observes 
two remarkable differences: In synchronized traffic the flow 
has a local minimum for a density of $30~$veh/km and reaches a 
second maximum for a density of $40~$veh/km. The origin of this
structure lies in the stochastic acceleration of cars which 
reduces considerably the probability to form a jam. A second 
important feature is the complex structure  of the fundamental 
in the presence of jams. For very high global densities one 
observes all three traffic states at the same time and no 
strict phase separation as, e.g., in case of  the VDR model. 
\begin{figure}
\begin{center}
\vspace{0.3cm}
\includegraphics[width=0.9\linewidth]{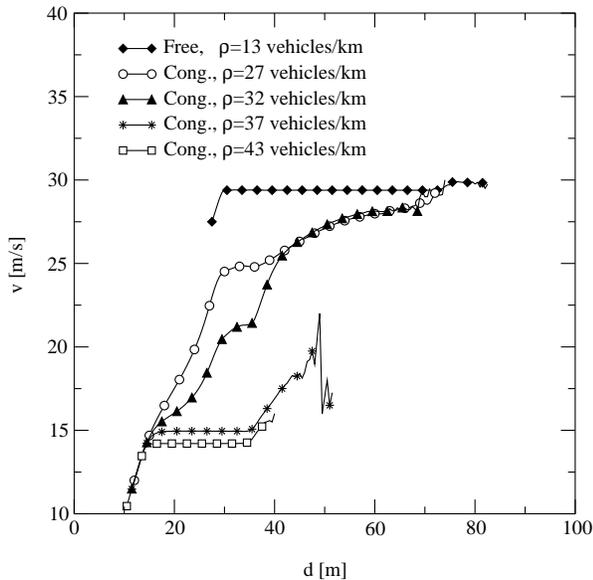}
\vspace{0.5cm}
\caption{OV function in free flow and congested traffic of the KKW 
model for different densities. The same set of parameters as in 
Fig.~\ref{fig:kkwlocalfund} has been used.} 
\label{fig:kkwov}
\end{center}
\end{figure}

Measurements of the OV-function show that the microscopic structure of
the model differs from the empirical findings. In free flow traffic
small headways are almost not observed and the maximum speed is
reached at larger distances than in real traffic. This indicates that,
compared to real traffic, the repulsive part of the car-car
interactions is overemphasized. While the differences between
empirical data and model results might be reduced for a different set
of model parameters, the model results for synchronized traffic differ
even qualitatively. In real traffic one observes for a given density a
crossover from a density independent form of the OV-function at small
distances to an asymptotic velocity for larger distances (see
Fig.~\ref{ov_emp}) which depends on the density.  This is not
reproduced by the KKW model, where a distance independent average
velocity is observed only in a narrow range of spatial headways, if it
is observed at all.

The comparison between empirical and simulation results of the 
time-headway distribution indicates that the model largely 
fails to reproduce the empirical results obtained for free flow
traffic. This is, as discussed before, partly a results of 
the simplified setup we used for our simulations. A much 
better agreement would be obtained if we consider a realistic 
distribution of maximal velocities. But even in this case 
one is left with a problem. The lack of velocity-anticipation leads to a
sharp cut-off  
of the time-headway distributions for times less than one 
unit of time, i.e. $1~\rm{s}$. Although the position of the 
cut-off can be tuned by varying the temporal discretization,
it must be noted that this still does not lead to the right 
functional form, as the maximum of the time-headway distribution 
is located at the minimal observed time-headway. This again 
confirms  necessity of velocity anticipation for the 
 reproduction of the  empirical findings at short time-headways.

Summarizing the CA model introduced by Kerner, Klenov and Wolf reveals
three distinguishable traffic states, as  observed in 
empirical studies. The reproduction of the empirical time-headway 
distribution and fundamental diagram is partly satisfying and 
could be easily improved by the introduction of velocity 
anticipation. The most important differences between empirical 
findings and model results concern the OV-function. 
This indicates that the microscopic structure of the model 
states does not match the real structure of highway traffic. 
We also believe that this disagreement is due to the very nature of the
car-car interactions in the KKW model and cannot be resolved by a better
choice of the model parameters.


\section{Comparison of the fundamental diagrams}
\label{global}

The comparison of the models presented so far is based on local
measurements of inductive loops. Therefore, the model parameters have
been chosen in order to allow the best possible accordance with the 
empirical setup. One of the main disadvantages of local measurements
is that the detected values of the flow and the velocity strongly fluctuate
whereas density cannot even be defined locally in a strict sense. However,
in traffic flow simulations it is possible to get averaged
quantities that are representative for a given density. Therefore, 
in this section global measurements of the flow and the density
of the various models are given for a typical set of parameters in
order to demonstrate the 
characteristics of the approaches. However, since density can be
calculated exactly, the distinction between the traffic states is omitted.

\begin{figure}[hbt]
\begin{center}
\includegraphics[width=0.9\linewidth]{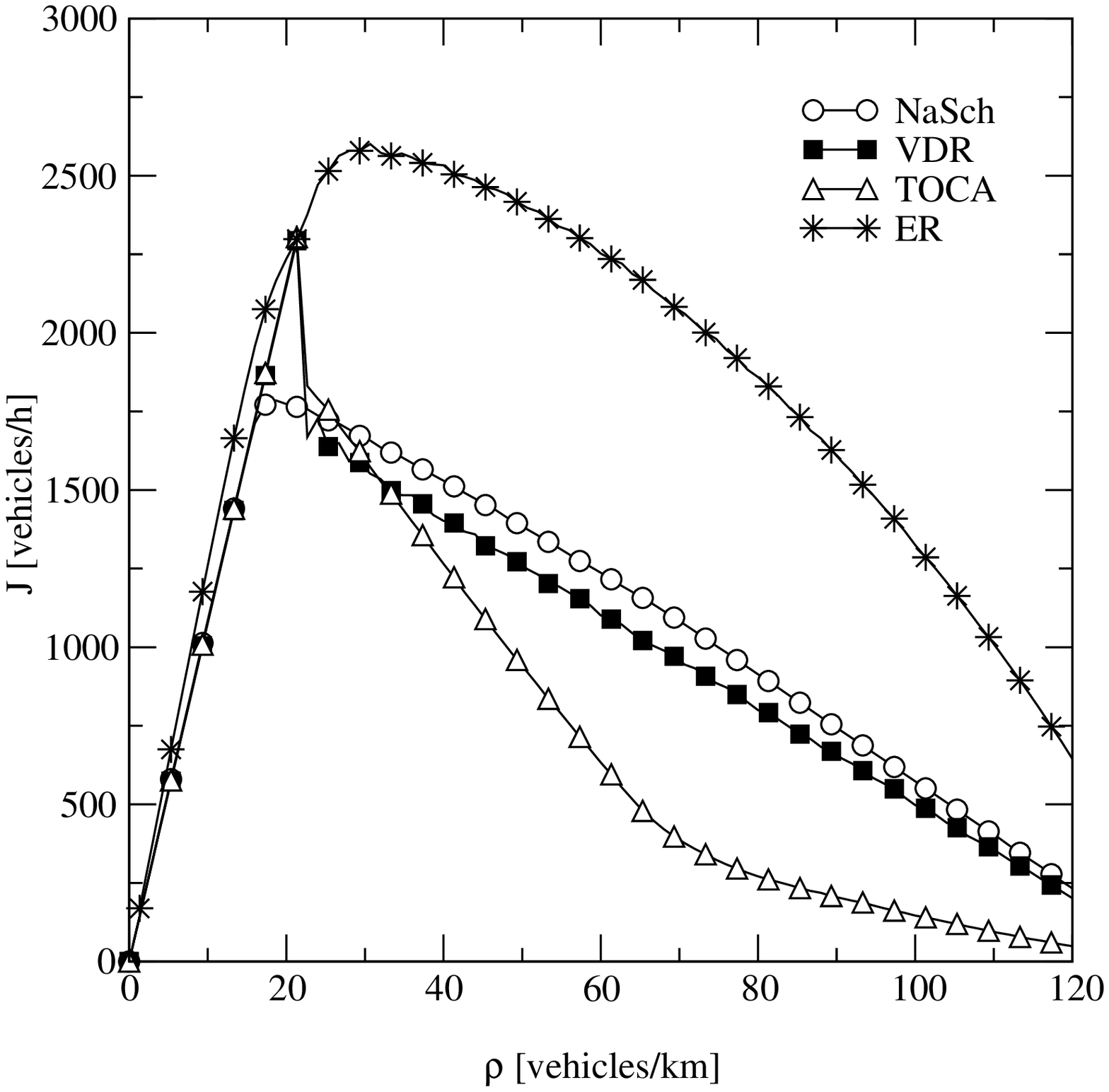}
\caption{Comparison of the global fundamental diagram of the NaSch
model, the VDR model, the TOCA, and the ER model for typical parameter values
and a homogeneous initialization.}
\label{global1}
\end{center}
\end{figure}

Density, flow and velocity can be measured globally in the following way:
The density $\rho_{\rm{global}}$ can directly be obtained by counting
the number $N$ of vehicles on a highway section of length $L$ via
\begin{equation} \rho_{\rm{global}} = \frac{N}{L}.\end{equation}
The average velocity $v_{global}$ is then defined as 
\begin{equation} v_{\rm{global}} = \frac{1}{N} \sum_{n=1}^{N} v_n
\end{equation} with the velocity $v_n$ of vehicle $n$.
Again, the hydrodynamical relation allows the calculation of the flow 
\begin{equation} 
J_{\rm{global}} = \rho_{\rm{global}} v_{\rm{global}} 
= \frac{1}{L} \sum_{n=1}^{N} v_n.
\end{equation}

\begin{figure}[hbt]
\begin{center}
\vspace{0.3cm}
\includegraphics[width=0.9\linewidth]{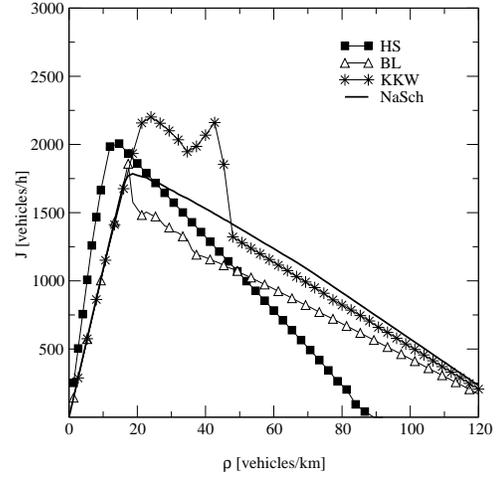}
\vspace{0.5cm}
\caption{Comparison of the global fundamental diagrams of the HS model, 
the BL model, and the KKW model with that of the
NaSch model for typical parameter values and 
a homogeneous initialization.}
\label{global2}
\end{center}
\end{figure}

A typical fundamental diagram consists of a linear free flow
branch that intersects with an almost linear congested branch.
As one can see in Fig.~\ref{global1} and Fig.~\ref{global2} nearly all
discussed models are able to reproduce this basic characteristics. The
fundamental diagram of the HS model, however, exhibits two distinct maxima.
The first maximum is simply given by the transition
from free flow to congested traffic. The second maximum is a
consequence of the chosen OV-curve. Since vehicles with $3 \le
d \le 5$ have to drive with a velocity of $2$, the flow
increases linearly for densities well
above a certain density until the average gap is smaller than $3$ cells.
Moreover, due to the OV-curve the vehicles behave deterministically
and choose their velocity according to the gap, i.e.\ $v=d$. 
As a result of a nearly uniform gap distribution, effectively speed limits 
are applied for certain density intervals which are reflected
by the occurrence of different slopes in the congested branch of the 
fundamental diagram. 

This behavior is typical for models with modified distance rules
and can also be found in the ER model. 
Since the choice of the gap-velocity matrix in
the ER model leads to speed limits for different density regimes, the
free flow branch shows two different
slopes like in the local measurements. Even more severe is
the lack of a distinct maximum 
in the fundamental diagram. This is a consequence of the ordered sequential
update of the ER model. It is possible that jams can also move in
downstream direction, thus leading to many small jams with a large
flow. 

Measurements of empirical data have revealed that the outflow from a
jam is reduced considerably compared to the maximum possible flow. As
a result, metastable free flow states exist and hysteresis effects
can be observed in the fundamental diagram \cite{kerner_prl79}. 

Obviously, this is the case for the VDR model, the TOCA model as well
as for the BL and KKW models while the maximum
possible flow of the NaSch model is as large as the outflow from a
jam. 

Since the
deceleration probability in the VDR model was chosen very small
($p_{{\rm dec}} = 0.01$) the stability of the homogeneous branch of the
fundamental diagram is very large. In contrast, once a jam has formed
above a certain threshold density the large deceleration probability
for the vehicles at rest is responsible for the reduced outflow from a
jam. As a result, the system is phase separated into a region
of free flow and a compact moving jam. The capacity drop can simply be tuned
by varying the 
difference between the two deceleration parameters.
In analogy to the VDR model, in the TOCA model only vehicles with $v
\le d$ decelerate with the probability $p$. Thus, vehicles driving
with $v_{\rm{max}}$ and $d > v$ lead to a stable high flow branch in
the fundamental diagram up to a density of $\frac{1}{v_{\rm{max}}+1}$.
However, the congested regime of the TOCA
model reveals the existence of two different slopes in the fundamental
diagram. For densities larger than $1/2$ vehicles have on average a
gap of less than one cell. Since the vehicles decelerate with a large
probability, but do accelerate with a rate smaller than one, the system
now contains only one large jam whose width is comparable to the 
system size.

\begin{figure}[hbt]
\begin{center}
\vspace{0.3cm}
\includegraphics[width=0.9\linewidth]{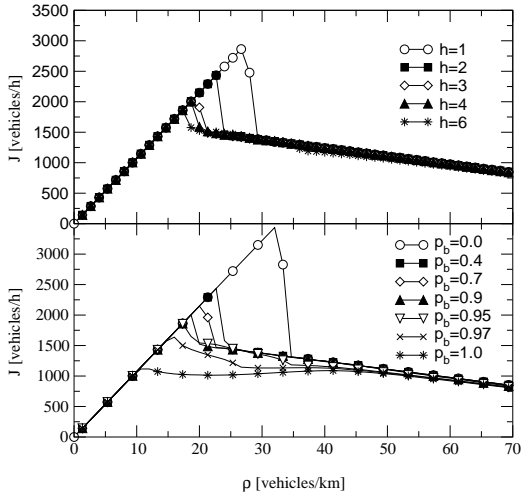}
\vspace{0.5cm}
\caption{Fundamental diagram for different  horizons
  $h$ (top) and for different  $p_{b}$ (bottom). All simulations
have been performed with an homogeneous initialization.} 
\label{carnaschglobalxp}
\end{center}
\end{figure}

Like in the VDR model, in the BL model the high flow states can simply
be controlled by the deceleration parameter $p_{0}$ for vehicles at
rest.  However, in the congested regime two distinct slopes of the
fundamental diagram become visible. The density at which the slope
changes and the shape of the fundamental diagram can be triggered by
the parameters $h$ and $p_{b}$ that determine the interaction between
vehicles with $d > v$.  In particular, the higher $h$ the smaller the
density $\rho_{\rm{max}}$ of the maximum flow
(Fig.~\ref{carnaschglobalxp} top).  For large $h$ the fundamental diagram
converges very fast, so that the fundamental diagram for values larger
than $h = 8$ are identical.  Moreover, even small values of $p_{b}$
have a strong influence on the flow.  The high flow branch of the
fundamental diagram (Fig.~\ref{carnaschglobalxp} bottom) and the density
$\rho_{\rm{max}}$ of maximum flow are reduced.  For large values the
congested branch of the fundamental diagram shows two different
slopes. The higher $p_{b}$, the smaller the density at which the slope
changes.


\subsection{Minimal model?}

The BL model  improves, compared to the other models we discussed
in this work, the agreement with  the  empirical 
data, especially  in the case of the  OV-curve.  Nevertheless, this is
only possible with    the  application of  a  variety   of new  update
rules. Therefore, it remains an  open question whether  this set of  update
rules can be reduced. 

\begin{figure}[hbt]
\begin{center}
\vspace{0.5cm}
\includegraphics[width=0.9\linewidth]{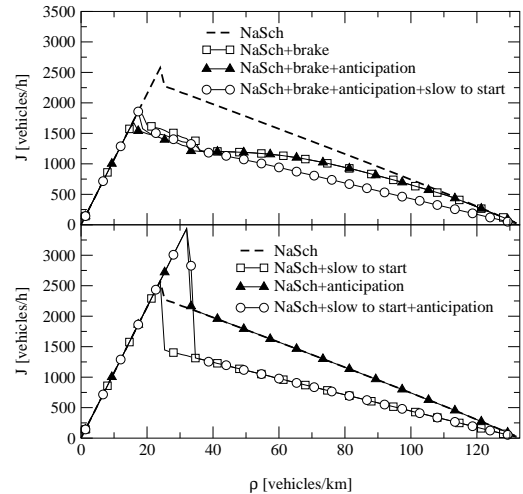}
\vspace{0.5cm}
\caption{Successive  extension of the  NaSch model  with brake
  lights (top) and without brake lights (bottom). Note that the system 
is initialized in an homogeneous state to generate also high flow states.}
\label{carnaschglobal}
\end{center}
\end{figure}

In the top part of Fig.~\ref{carnaschglobal} we successively dropped
the extensions of the model.  First, the slow-to-start rule has been
omitted. Without the slow-to-start rule the model lacks the ability of
a reduced outflow from a jam and the number of large compact jams is
reduced so that the flow increases.  As a further reduction of the
model, anticipation is switched off. This leads to a decrement of the
flow at densities larger than the density of maximum flow.  Now
headways smaller than the velocity are not possible, which manifests
in the OV-curve at small densities. For large densities the
anticipation of the predecessors velocity becomes more and more
difficult until anticipation is no longer applicable.  Therefore, the
differences between the curves with and without anticipation vanishes.

Applying the braking rule as the only extension leads to a
plateau-like fundamental diagram compared to the NaSch model.
Additionally, the flow is reduced dramatically.  It is the brake rule
that changes the shape of the fundamental diagram.

In bottom part of Fig.~\ref{carnaschglobal} the same successive reductions 
of the rules have been  applied to  the model without  braking  rule. 
Neither the anticipation, nor the slow-to-start rule applied as a
single extension or in combination are able to change the shape of the
fundamental diagram.

Considering the empirical fact that  small time-headways and a reduced
outflow from a jam exist,  the braking rule  is the only new extension
of the NaSch  model.  This new   rule  turns out  to be  crucial  for the
correct generation of the OV-curves and the occurrence of synchronized
traffic. 

So the set of rules chosen for the BL model is minimal in the
sense that all are needed to obtain a satisfactory agreement with
empirical data. We also believe that is essential to combine 
car-following like behavior and distance based rules. 
In case of the BL-model the velocity adjustment is event driven, i.e.\
the drivers react to braking cars in the upstream flow.
It is not excluded that the same can be
achieved with a different, but simpler set of rules. This is
highly desirable in order to reduce the complexity of the model
and the number of parameters. However, it is currently unclear
whether there is a similarly simple physical mechanism behind the formation
of synchronized traffic as it is behind the formation of wide jams.
For the latter, the reduction of the outflow from a jam below the
maximal flow is essential which can be easily achieved by any kind
of slow-to-start rule.


\section{Discussion}
\label{sec_disc}

The intention of our investigation was to single out the models which
are able to describe the empirically observed microscopic structure
of traffic flow correctly. It is well-known that many quite different
models exist which reproduce the macroscopic properties (e.g.\ global
fundamental diagrams or spontaneous jam formation) rather accurately
\cite{review,Helbing2000,nagatani}.
However, recently single-vehicle data have become available.
A thorough analysis of these data has allowed for a deeper understanding
of the microscopic properties which now should be incorporated into
the different modeling approaches.

We have suggested a test scenario based on the comparison
of computer simulations in a realistic setup with empirical data
obtained using stationary inductive loops. An important point is
that we have used only one fixed set of model parameters which
has been determined by comparison with empirical data, e.g.\
with the free-flow velocity. Therefore we are able to determine
whether a model is able to describe {\em all} traffic situations consistently
without the necessity to tune parameters according to the state.

Our focus was on cellular automata models \cite{review} and especially
variants of the Nagel-Schreckenberg model \cite{Nagel93,SSNI} which
can be considered as a minimal CA model for traffic flow. Our
comparison has revealed differences between the models on a
macroscopic scale which become even more pronounced on a microscopic
level of description.

We have seen that models with modified distance rules, like the ER and
the HS model, have problems on a macroscopic level. 
They are not able to produce a realistic (global) fundamental diagram
and it is difficult to make these models intrinsically crash-free.

The NaSch model, the VDR model, the TOCA model and the brake light
version of the NaSch model reproduce the fundamental diagram quite
well.  This is already sufficient for many applications. In urban
traffic, for example, the dynamics of the vehicles between two
intersections is predominantly determined by traffic lights. The
correct description of queues at cross-roads, therefore, only requires
the existence of two distinct traffic phases, namely free flow and
congested traffic.

More realistic applications of traffic flow simulations, e.g., that
allow the tracing of a jam, need a more detailed description of the
jamming mechanisms.  For the correct reproduction of the upstream
propagation of the downstream front of a jam it is necessary to reduce
the outflow from a jam and thus to facilitate metastable states.
Here, the VDR model, the TOCA model and the BL model allow the
existence of states with a flow considerably larger than the outflow
from a jam.

Differences between the models can be observed in the jam dynamics. 
While the road in the VDR model is separated into a region with
free flow and a compact jam that propagates upstream, the
peculiarities of the update rules of the TOCA model lead to a 
jam that covers the whole system.

Large compact jams appear also in the BL
model since the slow-to-start rule of the VDR-model is
included. However, brake lights are responsible for the generation 
of synchronized regions, i.e., regions of vehicles that are moving with
a small velocity but high flow.

This difference in the vehicle dynamics becomes most obvious in the
analysis of locally measured single-vehicle data.
On a microscopic level of description the main difficulty 
lies in the reproduction of small time-headways that
can be found at low densities in free flow and in the density
dependence of the velocity-distance relationship. This important
behavior of the OV-curve demonstrates that the driving strategy of a
vehicle depends strongly on the traffic state while the vehicles in
most modeling approaches adjust their velocity directly according
to their headway only, and therefore by the density.

As a first step towards a realistic modeling of highway traffic the 
interaction horizon of the original NaSch model has be to enhanced like 
in the TOCA, the ER and the HS model. 
However, in the TOCA and the HS model the cell length is not decreased
which is necessary in order to reproduce realistic acceleration
values. Therefore, the benefits of the increased interaction horizon
do not become visible.
Moreover, vehicles do react in a {\em static} manner
to a stimulus within the horizon. In particular, the velocity gap
matrix used in the ER model just leads to speed limits for certain
densities. 

A further step is the incorporation of the idea of event-driven
anticipation. In contrast to the static reaction described in the
previous paragraph it allows for a {\em dynamical response} that will
enable the vehicles to adjust their velocity to the actual traffic
situation regardless of the traffic density in front.  This idea is
realized in the BL and KKW models.  It turns out, that in case of the
BL model only the introduction of brake lights, which allow the timely
adjustment of the velocity to the downstream speed and can propagate
in upstream direction, allows the reproduction of synchronized
traffic.  Of course, there might alternative ways to model
synchronized traffic, but we believe that long-ranged event-driven
interactions between the vehicles are essential.

The use of an effective gap by means of velocity 
anticipation reduces velocity fluctuations in free
flow and leads to platoons of vehicles driving bumper-to-bumper.
It is also worth mentioning that this effect is of special importance 
in multi-lane traffic as shown in \cite{wolfgang}.

We have seen that the BL model allows to overcome the problems in the
reproduction of synchronized traffic encountered in the other modeling
approaches. It reproduces qualitatively the observed behavior.  Even
the quantitative agreement is in most cases very good although the
test scenario has neglected effects like disorder (different vehicle
and driver types) and boundary conditions. In contrast, in most other
approaches the discrepancies between empirics and model behavior can
already be seen on a qualitative level.  In particular, in the
simulations of the BL model three qualitatively different microscopic
traffic states are observed in accordance with the empirical results.
The deviations of the simulation results are mainly due to simple
discretization artifacts which do not reduce the reliability of the
simulation results. We also want to stress the fact that the agreement
is on a microscopic level.  This improved realism of the BL model
leads to a larger complexity of the approach compared to other models
of this type.  Nevertheless, due to the discreteness and the local
car-car interactions, very efficient implementations should still be
possible. Moreover, the adjustable parameter of the model can be
directly related to empirical quantities.  The detailed description of
the microscopic dynamics will also lead to a better agreement of
simulations with respect to empirical data for macroscopic quantities,
e.g., jam-size distributions.  Therefore we believe that this approach
should allow for more realistic micro-simulations of highway networks.

Our results show that the CA models for highway traffic have reached a
very high degree of realism. The most complete description of the
empirical findings is by means of the BL model. This is not surprising
since the model has been designed in order to reproduce data of local
measurements.  But anyhow it is important to know which aspects of
real traffic are described by a certain model, because in the end the
aspired accordance of a model with empirical observations strongly
depends on the goal of the particular application.  So it is useful to
use oversimplified model approaches in order to concentrate on
particular aspects of traffic flow phenomena~\cite{Nagel96b}.

Finally we want to emphasize that the results obtained from modelling
approaches also help to improve our understanding of the general
principles of traffic flow. We have seen the the complexity of
human behavior becomes more important if one wants to reproduce
its properties more accurately. In the simplest case only the
accident-avoidence is sufficient to reproduce the basic properties,
like free-flow and jammed phases. For synchronized traffic, however,
this is not sufficient. Here the results indicate that the dependence
of the driving-strategy on the traffic state becomes essential. Drivers
do not only want to avoid crashes, but also drive comfortably, e.g.\
by avoiding unnecessary large acceleration or deceleration.
This has been emphasized in \cite{knospe2002} and is implemented
in slightly different form in the BL and KKW models.

The next step would be the inclusion of other modelling approaches,
not only cellular automata models. Using a different test scenario
this has recently been done by Brockfeld and Wagner \cite{brockfeld}.
They have compared travel-time for various models (e.g.\ NaSch, VDR
and OVM) with empirical data. Using methods from optimization
theory to determine the best parameters it was surprisingly found
that all models produce similar results that are not in good
agreement with the data. The reason for this is not understood.
However, the performance of more sophisticated models (like BL and KKW)
has not been investigated in \cite{brockfeld}.\\[0.3cm]

{\em Acknowledgement:}\\ 

The authors have benefited from discussions with   
R.~Barlovi\'{c}, S.\ Grabolus, D.\ Helbing, T.~Huisinga, L.~Neubert, 
C.\ R\"ossel, and D.E.\ Wolf.  
L.~Santen acknowledges  support from the Deutsche 
Forschungsgemeinschaft under Grant No. SA864/2-1.
We also thank the Ministry of Economics
and Small Businesses, Technology and Transport
of  North-Rhine Westfalia   as  well  as to  the  Federal  Ministry of
Education  and Research of  Germany  for financial support (the latter
within the BMBF project ``SANDY'').


\begin{appendix}

\section{Continuous limit of the NaSch model}
\label{a}

The adjustment of the acceleration of the vehicles in the original
NaSch model to empirical values (that are about
$1~m/s^{2}$~\cite{ite}) requires the decrement of the length of a
cell (see also \cite{krauss,sg,sgas}). 

\begin{figure}[hbt]
\begin{center}
\includegraphics[width=0.9\linewidth]{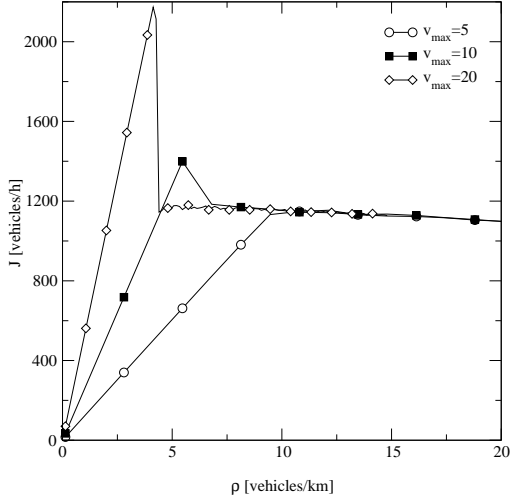}
\caption{NaSch model  with different
$v_{\text{max}}$ and homogeneous initialization.} 
\label{nasch_vinfty}
\end{center}
\end{figure}

This, however, entails an increment of the maximum possible
velocity for a given fixed absolute value of $\vmax$ (about
$100~km/h$ throughout this paper).  

\begin{figure}[hbt]
\begin{center}
\includegraphics[width=0.9\linewidth]{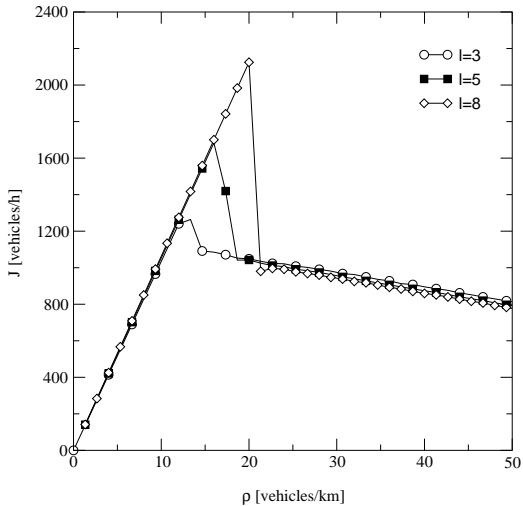}
\caption{Fundamental diagram  of the NaSch model  with
  $p_{\rm{dec}}    =    0.5$   for  different   discretizations    and
homogeneous   initialization.} 
\label{naschl}
\end{center}
\end{figure}

It turns out, that already the increment of the number of states a vehicle
is allowed to adopt leads to hysteresis effects of the flow. In particular, the
flow can be enhanced in a certain density regime by initializing
homogeneously the vehicles on the 
lattice compared to a pure random initial setup. As a result, in the 
limit $v_{\text{max}}\rightarrow \infty$~\cite{kertesz,huang,sg,sgas} 
the system exhibits metastable states (Fig.~\ref{nasch_vinfty}) 
with a flow increasing proportional to $v_{\text{max}}$, but with 
a rapidly decreasing lifetime.

Unfortunately, increasing only $\vmax$ leads to a significant
decrement of the density of maximum flow. 
Thus, in order to keep the maximum velocity fixed, the limit 
$\vmax \rightarrow \infty$ with $\vmax/l =$
const.~with the length $l$ of a cell has to be considered. 
Fig.~\ref{naschl} shows fundamental diagrams for different
finer discretization. 
Since the acceleration step of a vehicle is decreased
considerably, velocity fluctuations and vehicle interactions in free
flow are reduced. 
A random initialization of the system does not allow
the high flow states so that hysteresis can be observed.
On one hand, with increasing deceleration probability $p_{\rm{dec}}$ the
stability of the homogeneous flow branch of the fundamental diagram
decreases, but on the other hand the capacity drop increases.

Unlike in the VDR model, the origin of the high flow states cannot be
traced back to a reduction of the outflow from a jam but to the
stability of the free flow state.

A system with length $l$ and deceleration probability $p_{\rm{dec}}$
behaves like 
a NaSch model with cell length $1$ and a considerably smaller
deceleration probability of about $p_{\rm{dec}}/l$. In contrast, in
the congested 
regime the influence of the cell length can be neglected and a
system with decreased cell length behaves analogously to the NaSch
model with the same deceleration probability, e.g., the dynamics of
the vehicles in the congested regime of the NaSch model is maintained
(unlike in the cruise control limit of the NaSch
model~\cite{cruisecontrol}   where cars that are 
driving    with    $\vmax$     have a    deceleration    probability
$p_{\rm{dec}}(\vmax) = 0$). 

\begin{figure}[hbt]
\begin{center}
\includegraphics[width=0.9\linewidth]{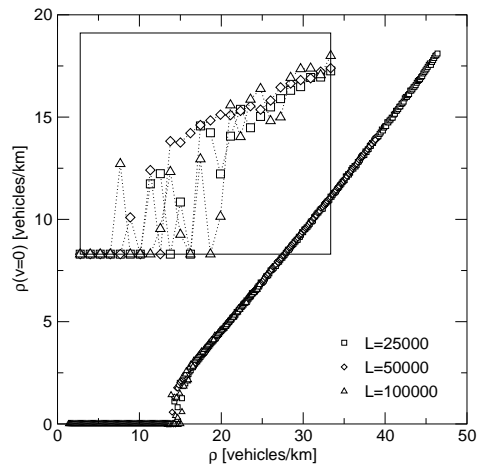}
\caption{Order-parameter $\eta$ for the NaSch model with a cell length 
of $1.5$~m and different system  sizes  $L$ for $p_{\rm{dec}} = 0.5$ and a
homogeneous initialization. The inset zooms into the transition region.} 
\label{neworder}
\end{center}
\end{figure}

For realistic traffic simulations it is important that the high flow
states are metastable for finite systems in the sense
that the probability for a perturbation that leads to a collapse of
the flow is only very small. Nevertheless, in the thermodynamic limit
the high flow states become unstable so that the homogeneous branch of
the fundamental diagram vanishes.

In order to study the phase transition we
introduced  an order  parameter  that exhibits   a qualitatively different
behavior within the  two phases. Because of  the  mass conservation in
the  NaSch  model with  periodic  boundary conditions we  observed the
density $\eta$ of jammed cars: 
\begin{equation}
 \eta = \frac{1}{L} \sum_{i=1}^{N}\delta_{v_{i},0}. 
\end{equation}
In  the   NaSch   model  $\eta$  decays exponentially   in the
vicinity of the transition~\cite{eisen} whereas a sharp drop occurs in
the VDR model~\cite{georg}. Due  to the finite braking  probability in
the NaSch model  cars with  zero  velocity do exist even  at densities
below the transition    density.   In contrast,  due  to  the    small
deceleration probability  $p_{\rm{dec}}$  in  the VDR model  one
macroscopic jam 
forms only at densities above the  transition density. With increasing
$p_{\rm{dec}}$ the transition smears out. 

Analogously to  the VDR model, the  order parameter of the NaSch model
with a finer  discretization (Fig.~\ref{neworder}) shows  a transition
from  zero  to a linear dependence of  the  density. With increasing
system size the  high-flow states become unstable  and the jump in the
order  parameter vanishes which demonstrates the metastability of the
high flow states. 


\section{Accidents in the HS-model: A static criterion}
\label{b}

In order to ensure collision-free motion in a model
with parallel update the condition
\begin{equation}
v_n(t+1) \leq d_n-1 +v_{n+1}(t+1)
\label{condition1}
\end{equation}
must always be fulfilled, 
i.e., the new velocity $v_n(t+1)$ of a car has to be smaller than
the number $d_n-1$ of empty cells in front plus the number $v_{n+1}(t+1)$
of cells the preceding car moves in the next time-step.
Eqn.~(\ref{condition1}) has to be complemented by the inequality
$0\leq v_n(t+1)$ that ensures that vehicles do not move backwards.

Consider now the case where the vehicle approaches the end of a jam,
i.e., the preceding car is standing and will not move in the next
time-step ($v_{n+1}(t+1)=0$). Using the acceleration rule eqn.~(\ref{accel}),
condition eqn.~(\ref{condition1}) can be rewritten as
\begin{equation}
v_n(t+1)+ \Big\lfloor \lambda [ V_{\rm{opt}}(d_n) - v_n(t+1) ]
\Big\rfloor \leq d_n-1 . 
\label{condition2}
\end{equation}
In order to be intrinsically free of collisions, condition
(\ref{condition2}) has to be satisfied for all $d$ and all $v$.
For $\lambda=1$ the inequality (\ref{condition2}) is always
satisfied if  $V_{\rm{opt}}(d_n) \leq d_n-1$. For general $\lambda$,
however, this is not the case.

This can easily be verified by initializing 
the system in a compact jam. In our simulations jams {\it always} occurred 
for global densities larger than $20$ veh/km when the first car arrived 
at the jam. This simulation result has to be discussed in the context 
of the empirical results of the jam dynamics. Empirically one observes 
quite often a jam surrounded by free flow traffic. This includes 
the fact that cars approach the upstream front of jams with a rather 
large velocity. Unfortunately for the HS model these kind of configurations 
lead to accidents, which is in sharp contrast to the real 
situation. 

But how does one have to choose $\lambda$ for a given OV function?
Using the inequalities $ x \geq \lfloor x \rfloor > x-1$ (for $x<0$)
one can derive sufficient conditions on the sensitivity parameter
$\lambda$ for the model to be {\em realistic} in the sense that no
collisions occur 
\begin{equation}
\lambda >
\max\left\{\frac{d-v-1}{V_{\rm{opt}}(d) - v}: v >  V_{\rm{opt}}(d)\right\},
\end{equation}
and vehicles do not move backwards
\begin{equation}
\lambda \leq \min\left\{\frac{v}{v-V_{\rm{opt}}(d)}: v >
V_{\rm{opt}}(d)\right\}. 
\end{equation}
We checked these two conditions for the OV-function given in~\cite{HeSch}.
It turns out that for the chosen $V_{\rm{opt}}$-function $\lambda = 1$ is 
the only possible choice. The upper limit for $\lambda$ holds for 
a quite general class of OV-functions, i.e., it is the upper limit 
for all OV-functions having $V_{\rm{opt}}=0$ for some value of the 
gap.

\end{appendix}



\begin{references} 

\bibitem{review}    D.~Chowdhury,   L.~Santen, and  A.~Schadschneider,
Physics Reports {\bf 329}, 199 (2000) and 
Curr.~Sci.~{\bf 77}, 411 (1999). 
\bibitem{Helbing2000} D.~Helbing, Rev.\ Mod.\ Phys.\ {\bf 73}, 1067 (2001).
\bibitem{nagatani} T.~Nagatani, Rep.\ Prog.\ Phys.\ {\bf 65}, 1331 (2002).
\bibitem{NWW} K. Nagel, P. Wagner, R. Woesler, Oper.\ Res.\ {\bf 51}, 
681 (2003)
\bibitem{herman} R.~Herman and  K.~Gardels, Sci.~Am.~{\bf 209} (6), 35
(1963). 
\bibitem{gazis61} D.C.~Gazis, R.~Herman, and R.W.~Rothery, Op.~Res.~9,
  454 (1961). 
\bibitem{gazis67} D.C.~Gazis, Science {\bf 157}, 273 (1967). 
\bibitem{gipps81} P.G.~Gipps, Transp.~Res.~{\bf 15}B, 105 (1981). 
\bibitem{rothery_trb} R.W.~Rothery,  in:   N.~Gartner,   C.J.~Messner,
  A.J.~Rathi  (eds.),  Transportation Research  Board  (TRB),  Special
  Report 165, {\em Traffic Flow Theory}, 2nd ed. (1998). 
\bibitem{Nagel93} K.~Nagel and M.~Schreckenberg,
J.~Physique I  {\bf 2}, 2221 (1992). 
\bibitem{SSNI} M.~Schreckenberg, A. Schadschneider, K.~Nagel, and N.\ Ito,
Phys.~Rev.~E {\bf 51}, 2939 (1995).
\bibitem{esser} J.~Esser and M.~Schreckenberg, 
Int.~J.~Mod.~Phys.~C{\bf 8}, 1025 (1997). 
\bibitem{krauss} S.\ Krauss, P.\ Wagner and C.\ Gawron,
Phys.~Rev.~E {\bf 54}, 3707 (1996), and {\bf 55}, 5597 (1997)
\bibitem{galilei} D.E.\ Wolf, Physica A {\bf 263}, 438 (1999).
\bibitem{robert}    R.~Barlovic,  L.~Santen,  A.~Schadschneider,   and
  M.~Schreckenberg, Eur.~Phys.~J. {\bf B5}, 793 (1998). 
\bibitem{brilon}  W.~Brilon and  N.~Wu (1999), 
in  {\it Traffic  and Mobility} ed. W.~Brilon, F.~Huber, M.~Schreckenberg,
and H.~Wallentowitz (Berlin: Springer) 
\bibitem{emmerich}  H.~Emmerich and E.~Rank,
Physica A {\bf 234}, 676 (1997).
\bibitem{HeSch}   D.~Helbing and   M.~Schreckenberg,
  Phys.~Rev.~E {\bf 59}, R2505 (1999). 
\bibitem{knospe2001} W.~Knospe, L.~Santen, A.~Schadschneider, and
M.~Schreckenberg, J.~Phys.~A{\bf 33}, L477 (2000).
\bibitem{knospe2002} W.~Knospe, L.~Santen, A.~Schadschneider, and
M.~Schreckenberg, Phys.~Rev.~E {\bf 65}, 015101 (2002).
\bibitem{kkw} B.S.~Kerner, S.L.\ Klenov, and D.E.\ Wolf, 
J.~Phys.~A{\bf 35}, 9971 (2002).
\bibitem{neubert}       L.~Neubert,    L.~Santen,   A.~Schadschneider,
and M.~Schreckenberg, Phys.~Rev.~E {\bf 60}, 6480 (1999). 
\bibitem{Tilch99TGF} B.~Tilch and D.~Helbing, In: {\em Traffic and Granular
Flow '99}, p.~333, D.~Helbing, H.~J.~Herrmann, M.~Schreckenberg, D.~E.~Wolf
(eds.) Springer (Heidelberg, 2000).
\bibitem{knospe2002b} W.~Knospe, L.~Santen, A.~Schadschneider, and
M.~Schreckenberg, Phys.\ Rev.\ E {\bf 65}, 056133 (2002).
\bibitem{mahnke} I.A.\ Lubashevsky, R. Mahnke, P.\ Wagner, and S.\ Kalenkov,
Phys.\ Rev.\ E {\bf 66}, 016117 (2002).
\bibitem{kernerNet} B.S.\ Kerner, Netw.\ Spatial Econo.~{\bf 1}, 35 (2001)
\bibitem{HTcoop} D.\ Helbing, M.\ Treiber, Coop.\ Tr@nsp.\ Dyn.\ {\bf 1}, 
paper 2 (2002)
\bibitem{hall} F.~L.~Hall, B.~L.~Allen,and M.~A.~Gunter, Transp.~Res.~A
{\bf 20A}, 197 (1986). 
\bibitem{kernerphysworld} B.~S.~Kerner, Phys.~World 8, {\bf 25} (1999). 
\bibitem{NeubertDiss} L.~Neubert, {\em Statistische Analyse von
Verkehrsdaten und die Modellierung von Verkehrsfluss mittels
zellularer Automaten}, Ph.D.~thesis, University Duisburg (2000).
\bibitem{kerner_prl81} B.S.~Kerner, 
Phys.~Rev.~Lett.~{\bf 81}, 3797 (1998). 
\bibitem{kerner96} B.S.~Kerner and H.~Rehborn, 
Phys.~Rev.~E {\bf 53}, R1297 (1996). 
\bibitem{bando} M.~Bando and K.~Hasebe, A.~Nakayama, A.~Shibata, and
Y.~Sugiyama, Phys.~Rev.~E {\bf 51} 1035 (1995).
\bibitem{Fukui96} M.~Fukui and Y.~Ishibashi, Phys.~Rev.~E {\bf 51},
2339 (1995).
\bibitem{ultra} K.\ Nishinari and D.\ Takahashi, J.\ Phys.\ A {\bf 32}, 
93 (1999)
\bibitem{schuetz} G.M. Sch\"utz, in:  {\it Phase Transitions
and Critical Phenomena}, Vol.19, eds. C. Domb and J.L. Lebowitz
(Academic Press, 2000)
\bibitem{debch} K.~Ghosh, A.~Majumdar, and D.~Chowdhury,
Phys.~Rev.~E {\bf 58}, 4012 (1998).
\bibitem{annalen}   A.~Schadschneider and M.~Schreckenberg, 
Ann.~Physik {\bf 6}, 541 (1997). 
\bibitem{tt} M.~Takayasu and H.~Takayasu, Fractals {\bf 1}, 860 (1993). 
\bibitem{appert} C. Appert, L.~Santen, Phys.~Rev.~Lett.~{\bf 86}, 2498 (2001)
\bibitem{robert_neu} R.\ Barlovic, T.\ Huisinga, A.\ Schadschneider, and
M.\ Schreckenberg, Phys.\ Rev.\ E {\bf 66}, 046113 (2002).
\bibitem{localcluster} B.S.~Kerner and P.~Konh\"auser:
Phys.~Rev.~E {\bf 48}, R2335 (1993).
\bibitem{Rajewsky} N.\ Rajewsky, L.\ Santen, A.\ Schadschneider, and
M. Schreckenberg, J.~Stat.\ Phys.~{\bf 92}, 151 (1998).
\bibitem{HT} D.\ Helbing and B.\ Tilch, Phys.\ Rev.\ E {\bf 58}, 133 (1998).
\bibitem{Sasoh} A.\ Sasoh, J.\ Phys.\ Soc.\ Jpn.\ {\bf 70}, 3161 (2001).
\bibitem{barret} C.~L.~Barrett, and M.~Wolinsky, 
in {\it Traffic and Granular Flow} ed. D.~E.~Wolf, M.~Schreckenberg 
and A.~Bachem (Singapore: World Scientific), p.169. 
\bibitem{wolfgang}  W.~Knospe,   L.~Santen,     A.~Schadschneider, and
  M.~Schreckenberg, Physica A {\bf 265}, 614 (1999). 
\bibitem{bremslichtpaper}  M.~Goldbach,   A.~Eidmann, and A.~Kittel,
Phys.~Rev.~E {\bf61}, R1239 (2000). 
\bibitem{HermanRothery} R.~Herman  and R.W.~Rothery, Car Following and
  Steady-State Flow, Proceedings of the 2nd International Symposium on
  the Theory of   Traffic Flow, Ed.~W.~Leutzbach   and P.~Baron, Bonn,
  Germany, 1965. 
\bibitem{ite}  Institute of  Transportation Engineers (1992).  Traffic
  Engineering Handbook, Washington, DC. 
\bibitem{george} H.P.~George,  
{\em Measurement  and Evaluation  of Traffic   Congestion}, 
Bureau of Highway Traffic, Yale University 1961, 43-68. 
\bibitem{miller} A.J.~Miller, 
Jrl.\ Roy.~Statist.~Soc.\ B {\bf 23} (1961) 1. 
\bibitem{schlums}  J.~Schlums, 
{\em Untersuchungen des Verkehrsablaufes auf Landstra{\ss}en}, 
Eigenverlag, Hannover (1955). 
\bibitem{hcm} Highway  Capacity  Manual, U.S.~Department  of Commerce,
  Bureau  of Public  Roads,  Washington  D.C.~1965,  HRB Spec.~Rep.~87
  (1965) 
\bibitem{edie}  L.C.~Edie and R.S.~Foote,   
  Proc.~HRB {\bf 37}, 334  (1958). 
\bibitem{pfefer} R.C.~Pfefer, 
Transportation Engineering Journal of American Society of
  Civil Engineers, Vol.~102, No.~TE4, 683-697 (November 1976). 
\bibitem{kerner_prl79} B.S.~Kerner and H.~Rehborn, 
Phys.~Rev.~Lett.~{\bf 79}, 4030 (1997). 
\bibitem{kk} B.S.~Kerner and S.L.\ Klenov, J.~Phys.~A{\bf 35}, L31 (2002).
\bibitem{kk2} B.S.~Kerner and S.L.\ Klenov, 
Phys.~Rev.~E  {\bf 68}, 036130 (2003).
\bibitem{Nagel96b} K.~Nagel, Phys.~Rev.~E {\bf 53} (5), 4655.
\bibitem{sg} S.\ Grabolus, {\em Numerische Untersuchungen zum
Nagel-Schreckenberg-Verkehrsmodell und dessen Varianten},
Diplomarbeit, Universit\"at zu K\"oln, Germany (2001).
\bibitem{sgas} S.\ Grabolus and A.\ Schadschneider, in preparation.
\bibitem{kertesz} M.~Sasv\'{a}ri and J.~Kert\'{e}sz, 
Phys.~Rev.~E  {\bf 56}, 4104 (1997). 
\bibitem{huang} Ding-wei Huang and  Chung-wei Tsai,  
Phys.~Rev.~E  {\bf 61}, 012101 (2001). 
\bibitem{cruisecontrol}   K.~Nagel  and M.~Paczuski,  
Phys.~Rev.~E {\bf 51}, 2909 (1995). 
\bibitem{eisen} B.~Eisenbl\"atter, L.~Santen,  A.~Schadschneider, and
  M.~Schreckenberg, Phys.~Rev.~E {\bf 57}, 2 (1998). 
\bibitem{georg} G.~Diedrich, {\em Numerische Untersuchung zur
Phasenseparation in Zellularautomaten f\"ur Strassenverkehr},
Diplomarbeit, Universit\"at zu K\"oln, Germany (1999).
\bibitem{jiang} R.\ Jiang and Q.-S.\ Wu, J.\ Phys.\ A {\bf 36}, 381 (2003)
\bibitem{krbalek} M.\ Krbalek and D.\ Helbing, 
Physica {\bf A333}, 370 (2004)
\bibitem{brockfeld} E.\ Brockfeld and P. Wagner,
in {\em Interface and Transport Dynamics --- Computational Modelling},
edited by H.\ Emmerich, B.\ Nestler, M.\ Schreckenberg (Springer 2003)
\end{references}
\end{document}